\documentclass[twocolumn]{aastex62}
\usepackage{natbib}
\graphicspath{{./}{pictures/}}

\received{}
\revised{}
\accepted{}
\submitjournal{ApJ}

\shorttitle{Tidal disruption events}
\shortauthors{Ryu et~al.}

\usepackage{graphicx}	
\usepackage{amsmath}	
\usepackage{amssymb}	
\usepackage{cases}
\usepackage{array,multirow}
\usepackage{upgreek}
\usepackage{textcomp}
\usepackage{hyperref}
\usepackage[referable]{threeparttablex}
\usepackage{booktabs, dcolumn}
\usepackage{xspace}

\usepackage{ulem}

\newcommand{\harm}{{\sc Harm3d}\xspace}   
\newcommand{\mesa}{{\small MESA}\xspace}

\newcommand*{\doverline}[1]{\overline{\overline{#1}}}

\newcommand{\beq}{\begin{equation}}
\newcommand{\eeq}{\end{equation}}
\newcommand{\simlt}{\mathrel{\hbox{\rlap{\hbox{\lower4pt\hbox{$\sim$}}}\hbox{$<$}}}}
\newcommand{\simgt}{\mathrel{\hbox{\rlap{\hbox{\lower4pt\hbox{$\sim$}}}\hbox{$>$}}}}

\newcommand{\erg}{\;\mathrm{erg}}
\newcommand{\s}{\;\mathrm{s}}

\newcommand{\Msol}{\;\mathrm{M}_{\odot}}
\newcommand{\Rsol}{\;\mathrm{R}_{\odot}}
\newcommand{\gram}{\;\mathrm{g}}

\newcommand{\Gyr}{\;\mathrm{Gyr}}

\newcommand{\K}{\;\mathrm{K}}

\newcommand{\rtidal}{\;r_{\rm t}}
\newcommand{\rg}{r_{\rm g}}
\newcommand{\physrad}{\mathcal{R}_{\rm t}}
\def\apjl{ApJL}
\def\apj{ApJ}
\def\mnras{M.N.R.A.S.}
\def\aap{A\&A}
\def\nat{Nat.}

\def\araa{Ann. Rev. A\&A}
\def\pasp{PASP}
\def\apjs{ApJ Supp.}

\def\prd{prd}

\defcitealias{Ryu1+2019}{Paper 1}
\defcitealias{Ryu2+2019}{Paper 2}
\defcitealias{Ryu3+2019}{Paper 3}
\defcitealias{Ryu4+2019}{Paper 4}
\begin{document}

\title{Tidal disruptions of main sequence stars - II. Simulation methodology and stellar mass dependence of the character of full tidal disruptions}

\correspondingauthor{Taeho Ryu}
\email{tryu2@jhu.edu}

\author[0000-0002-0786-7307]{Taeho Ryu}
\affil{Physics and Astronomy Department, Johns Hopkins University, Baltimore, MD 21218, USA}

\author{Julian Krolik}
\affiliation{Physics and Astronomy Department, Johns Hopkins University, Baltimore, MD 21218, USA}
\author{Tsvi Piran}
\affiliation{Racah Institute of Physics, Hebrew University, Jerusalem 91904, Israel}

\author{Scott C. Noble}
\affiliation{Gravitational Astrophysics Laboratory, Goddard Space Flight Center, Greenbelt, MD 20771, USA}

\begin{abstract}

This is the second in a series of papers presenting the results of fully general relativistic simulations of stellar tidal disruptions in which the stars' initial states are realistic main-sequence models. In the first paper \citep{Ryu1+2019}, we gave an overview of this program and discussed the principal observational implications of our work. Here  we describe our calculational method, which includes a new method for calculating fully-relativistic stellar self-gravity, and provide details about the outcomes of full disruptions, focusing on the stellar mass dependence of the outcomes for a black hole of mass $10^{6}\rm{M}_{\odot}$. We consider eight different stellar masses, from $0.15~{\rm M}_\odot$ to $10~{\rm M}_\odot$. 

We find that, relative to the traditional order-of-magnitude estimate $r_{\rm t}$, the physical tidal radius of low-mass stars ($M_{\star} \lesssim 0.7~ {\rm M}_\odot$) is larger by tens of percent, while for high-mass stars ($M_{\star} \gtrsim1~ {\rm M}_\odot$) it is smaller by a factor 2--2.5.  The traditional estimate of the range of energies found in the debris is $\approx 1.4\times$ too large for low-mass stars, but is a factor $\sim 2$ too small for high-mass stars; in addition, the energy distribution for high-mass stars has significant wings.  For all stars undergoing tidal encounters, we find that mass-loss continues for many stellar vibration times because
the black hole's tidal gravity competes with the instantaneous stellar gravity at the star's surface 
until the star has reached a distance from the black hole $\sim O(10)r_{\rm t}$.

\end{abstract}


\section{Introduction} \label{sec:intro}

Observations suggest that almost every massive galaxy hosts at least one supermassive black hole (SMBH) in its center \citep{KormendyHo2013}. As stars in a galaxy's core interact gravitationally, some stars' orbits can be perturbed in a way that places them on nearly radial orbits. If they  approach the central BH sufficiently close, these stars are tidally disrupted and lose some fraction of  their mass. Roughly half the stellar debris is bound and returns back to the BH, while the other half is expelled outward at $\sim 5000$--10,000~km/s, producing a luminous flare. A few dozen candidate tidal disruption events (TDEs) have been identified \citep{Komossa2015,UV2018}, and the number is expected to grow with detections by the ongoing optical time-domain survey \citep[e.g., ZTF\footnote{The Zwicky Transient Facility}:][]{Graham+2019} as well as future surveys (e.g., eROSITA\footnote{Extended Roentgen Survey with an Imaging Telescope Array} All-Sky Survey: \citealt{Merloni+2012}, and LSST\footnote{The Large Synoptic Survey Telescope}: \citealt{LSST2009}). 

This paper is the second in a series of closely-related papers in which we seek to determine quantitatively the principal characteristics of main sequence stars disrupted by supermassive black holes.  To do so, we employ fully general relativistic dynamics operating on stars with realistic internal structures.  Here we focus on two aspects of our program: our calculational methodology, and how TDE properties depend on stellar mass\footnote{\citetalias{Ryu1+2019} gives an overview and the main observational implications of our results; \citetalias{Ryu3+2019} discusses partial disruptions in detail; \citetalias{Ryu4+2019} describes how relativistic effects lead to black hole mass-dependence.}.  For the latter purpose, we consider encounters between a $10^6\Msol$ black hole and main-sequence (MS) stars with different masses spanning the range $0.15\Msol \leq M_{\star} \leq 10\Msol$, whose initial state is taken from MESA models.

A detailed description of our methods is given in Section~\ref{sec:numericalsetup}, including discussion of: the code we use (Section~\ref{sec:Numericalmethod}); computational domain setup (Section~\ref{subsec:computationaldomain}); spacetime geometry, tidal stresses, and self-gravity (Section \ref{sec:Spacetime}); and our stellar models (Section~\ref{sec:stellarmodel}). Section~\ref{sec:procedure} presents our specific procedures: preparation of initial conditions (Section~\ref{sec:initial}); definition of stellar trajectories (Section~\ref{sec:stellartrajectories}); Our criteria for distinguishing partial disruptions from complete ones  (Section~\ref{subsec:disruptioncondition}).
In Section \ref{sec:results}, we give a detailed description of our results for full disruptions. In particular, we present the physical tidal radius $\mathcal{R}_{\rm t}$, defined as the maximum pericenter for full disruption (Section~\ref{sec:physrtidal}), and we discuss the duration of tidal disruption (Section~\ref{subsub:duration}), the debris energy and angular momentum distributions, and the fallback rate of the debris (both in Section \ref{subsub:dmde_ful}).
In Section~\ref{subsec:analyticmodel}, we show how the semi-analytic models introduced in \citetalias{Ryu1+2019} (predicting $\physrad$ from stellar central density, and the functional relation between remnant mass and pericenter for partial disruptions) were derived.
In Section~\ref{sec:discussion}, we compare our results for the physical tidal radius $\mathcal{R}_{\rm t}$ (Section~\ref{subsub:comparison_rt}) and the characteristic debris energy width (Section~\ref{subsub:energydebris}) with those found in other studies. Lastly, we summarize our results in Section~\ref{sec:summary}.

Throughout this paper, symbols with the subscript $\star$, such as $\tau_{\star}$ (stellar vibration time, defined in Section~\ref{sec:stellarmodel}), $R_{\star}$ (stellar radius) and $M_{\star}$ (stellar mass), always pertain to the initial model star. All masses are measured in units of $M_\odot$ and all stellar radii in units of $R_\odot$.

\section{Simulation Methodology}\label{sec:numericalsetup}

\subsection{Numerical Method}
\label{sec:Numericalmethod}

We use the intrinsically conservative general relativistic magneto-hydrodynamics (GRMHD) code \harm 
\citep{Noble+2009}.
The code is an extended version of the 2D GRMHD {\small HARM}   \citep{Gammie+2003}. Like {\small HARM}, we adopt the Lax-Friedrichs numerical flux formula, but use a parabolic interpolation method \citep{ColellaWoodward1984} with a monotonized central-differenced slope limiter instead of {\small HARM}'s linear methods. \harm  has been used for studying many problems in BH physics, including energy production in accretion onto Kerr black holes \citep{Noble+2009,Noble2010,Schnittman+2013}, accretion flow from a stellar tidal disruption \citep[e.g.,][]{Shiokawa+2015}, accretion onto binary black holes \citep[e.g.][]{Noble+2012,Ascoli+2018}, and the X-ray spectra of stellar-mass black holes \citep{Kinch+2019}.

The equations solved in our application of \harm\  are $\nabla_\mu T^\mu_\nu = 0$ and $\nabla_\mu \rho u^\mu =0$, where the stress-energy tensor $T^\mu_\nu = \rho h u^\mu u_\nu - p g^\mu_\nu$, $\rho$ is the proper rest-mass density, $h$ is proper enthalpy $1 + p/\rho$, $p$ is the proper pressure, and $u^\mu$ is the fluid 4-velocity.
For the work presented here, the magnetic field evolution, normally a part of a \harm simulation, is turned off, as only non-magnetized stars are considered.

We assume an adiabatic equation of state with an adiabatic index $\gamma = 5/3$.  In real stars, the effective adiabatic index can differ from $5/3$, and \mesa~ employs equation of state tables constructed on the basis of quantum statistical calculations by \citet{RogersNayfonov2002} and \citet{Saumon+1995}.
However, the resulting effective adiabatic index wherever $T\geq 10^{5}\K$, i.e., in the bulk of the stellar mass, is $\simeq 5/3$.  In the course of the TDE, both the density and temperature of the stellar material decrease.
The only physical effect in the debris that might alter the adiabatic index is ionization state change, particularly where the temperature is low  enough for H to recombine.  Because, for the great majority of the stellar mass, H recombination takes place outside our simulation domain, $\gamma = 5/3$ is a well-justified approximation.

In any code adopting a conservative integration scheme, the transformation between the conserved quantities and the so-called primitive variables is performed at least once each time step per computational cell. In a conservative GRMHD code, the transformation between the two sets of variables is not straightforward because simple analytic relations between the two sets do not exist. In our study, we numerically recover the primitive variables from the conserved variables assuming conservation of momentum (spatial components of the conservation law of the stress-energy tensor, or specifically Equation 27 in \citealt{Noble+2006}) and entropy (Equation 19 in \citealt{Noble+2009}). Maintaining constant entropy means that all shocks are radiative to the degree imposed by this condition, but we do not see any significant shocks in our simulations in any case.

\begin{figure*}
	\centering
\includegraphics[width=15.4cm]{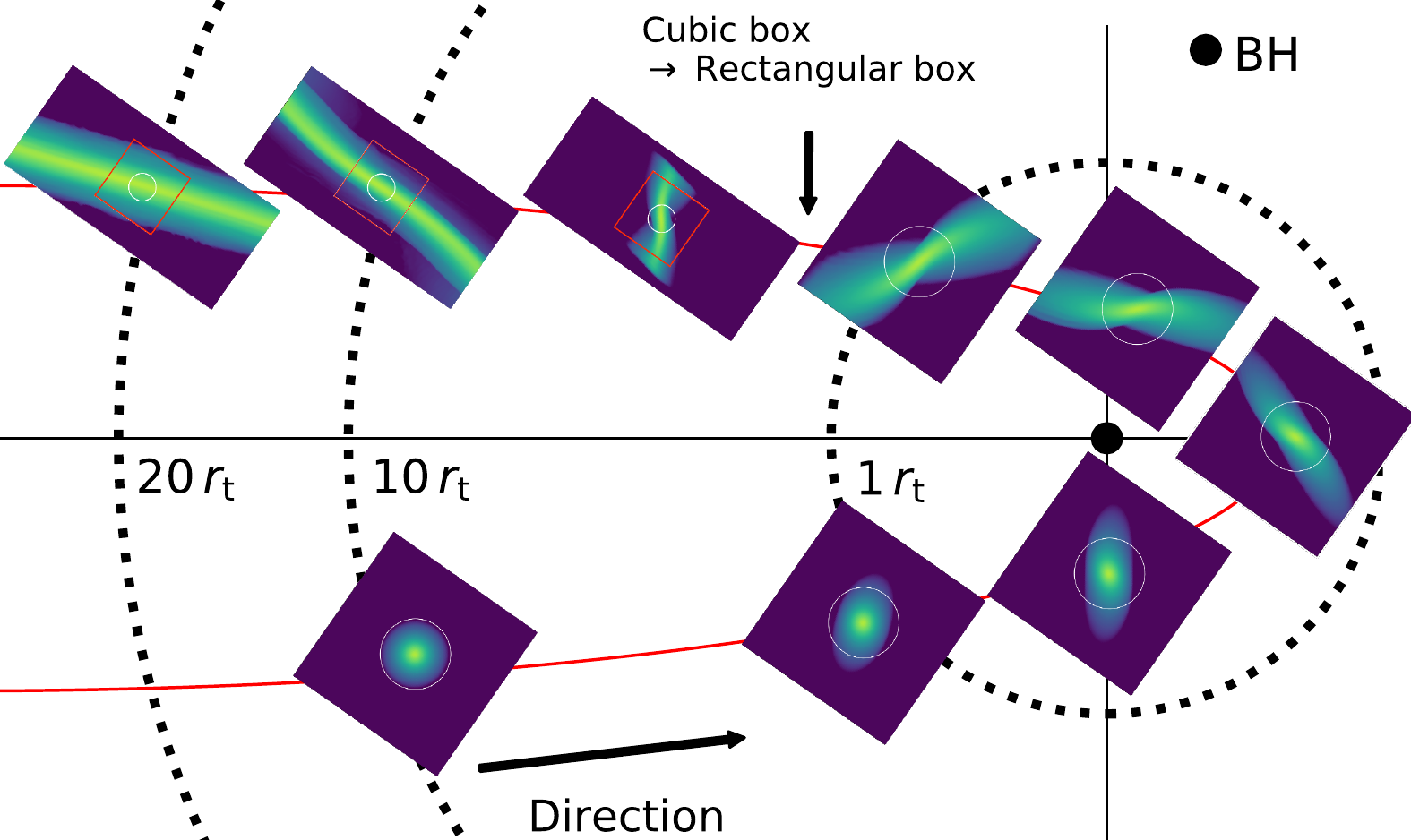}
\caption{Successive moments in a full TDE. The red line indicates the star's orbit around the black hole (black circle). Each inset figure presents a snapshot of the density distribution in the orbital plane within our simulation box.
The white circle in each snapshot shows the initial stellar radius. Partway through the event, we replace the cubic box with a rectangular box; we  draw a red square in the rectangular boxes to show the position and size of the original cubic box. Note that the rectangular boxes are not drawn to the same scale as the cubic boxes, and the dotted curves marking $\rtidal$, $10\rtidal$ and $20\rtidal$ are likewise not drawn to scale.}
\label{fig:overview}
\end{figure*}

\subsection{Computational domain}
\label{subsec:computationaldomain}
Our computational domain is a rectangular box of fixed orientation that moves with the star.  Midway through the simulation, we  change the spatial size and shape of the box
to accommodate the changing shape of the star and the debris.  We use a cubic box until the star's Boyer-Lindquist radial coordinate $r$ reaches 2--$4\rtidal$ from the BH as it moves away from  pericenter passage (here $\rtidal$ refers to the usual order-of-magnitude estimate for the tidal radius $\rtidal =\left(M_{\rm BH}/M_{\star}\right)^{1/3}R_{\star}$). 
At this point we replace it with an elongated rectangular box, larger in every dimension. We do so to ensure that the gas velocity is consistently supersonic outward at the box boundary.
In a small number of cases for which the pericenter $r_{\rm p}$ is well outside $\mathcal{R}_t$, tidal effects are so weak that replacing the cubic box is unnecessary.  The specific parameters of these boxes are:

\begin{enumerate}

\item \textit{Cubic domain: from onset ($r\simeq10\rtidal$ before pericenter passage) to $r\simeq 2-4\rtidal$ after pericenter passage.} \\
The sides of the cubic box are  $L_{x}=L_{y}=L_{z}=5 R_{\star}$. The resolution of the cubic box is $\approx25$ cells per $R_{\star}$.  The number of cells on each axis is $128$.\\
\item \textit{Extended elongated domain: from $r\gtrsim 2-4\rtidal$ after pericenter passage to the end of the simulation ($r \simeq (20-30)\rtidal$.) }\\
As the star is stretched by the tidal forces of the BH, the star becomes elongated primarily in one dimension. When the size in that direction becomes longer than the width of the cubic box, we increase the box in all dimensions, but more in the dimension most nearly parallel to the axis of debris extension. The  size of the larger box is $(L_{x},L_{y},L_{z}) = (17,~9,~14)\times R_{\star}$. For the larger rectangular domain, we coarsen the grid by a factor of 2 in all dimensions. \\

\item \textit{Orientation:} Because the direction of the debris extension is predictable, we start with a box rotated with respect to the semimajor axis of the orbit (as illustrated in Fig.~\ref{fig:overview}) so that the debris is extended along the $x$-axis of the grid when the box has traveled out to $r>10~r_{\rm t}$.  By this means, we can, in all cases, keep the angle between the $x$-axis and the stream to $< 25^\circ$ throughout the event.  \\

\end{enumerate}

We map the last snapshot of the cubic box onto the corresponding in the elongated domain in a manner ensuring that the total mass, momentum, and internal energy are conserved. The rest of the extended domain is filled with gas at the floor density.  Figure~\ref{fig:overview} schematically depicts how a star evolves in the comoving computational domain as it travels along an orbit, and how we change the computational box accordingly.

We also ran several simulations with a cubic box $2\times$ larger than the standard in all dimensions and a rectangular domain  $1.5\times$  larger than the standard size. We find no significant differences between runs with the different box sizes in terms of mass contained in the same volume around the domain origin and distinguishing between full and partial disruptions. We have also performed convergence tests
 with coarser and  with $1.5\times$ finer resolution and find no significant differences between those simulations and runs with our standard resolution (see Appendix \ref{appendix2} for more details about the convergence tests).

 We give all primitive variables zero gradient at boundaries. However, to ensure outflow, we set the normal component of the primitive fluid velocity in the ghost cells to be zero if the fluid motion is found to be inward.  The time-step is determined using a Courant number of 0.3.

\subsection{Spacetime geometry of the simulation: tidal gravity and self-gravity}
\label{sec:Spacetime}
\subsubsection{Tidal gravity and definition of the box frame}

All our simulations are carried out in a global Schwarzschild spacetime, but modified to include the star's self-gravity within the computational box.  To accomplish this, we proceed in a series of steps.  These begin by describing the Schwarzschild spacetime in terms of Cartesian coordinates with an origin at the black hole and oriented so that the $x$-axis is parallel to the orbital major axis.  We then transform this metric to the moving frame of the box by a coordinate transformation in which the time coordinate does not change.  The last step of this transformation is to rotate the spatial coordinate axes to align with the box sides.  We call the resulting coordinate system the ``box frame".  This procedure guarantees that the relativistic tidal gravity of the black hole is expressed exactly in the frame of the computational box.  Note that because we fix the time coordinate, this is {\it not} a Lorentz transformation.

\subsubsection{The self-gravity component $h_{\mu\nu}^{\rm sg}$}
\label{subsec:selfgravity}

The easiest way to combine stellar self-gravity with the background metric is to use a post-Newtonian approximation.  In this approximation, the total metric is
\begin{align}\label{eq:metric}
g_{\mu\nu} \simeq \tilde g_{\mu\nu} + h_{\mu\nu}^{\rm sg},
\end{align}
where $\tilde g_{\mu\nu}$ 
is the global Schwarzschild metric as it is represented in the box frame, and
\begin{align}\label{eq:postN}
h_{00}^{\rm sg} &= -2 \Phi_{\rm sg}c^{-2},\\
h_{0i}^{\rm sg}  =h_{i0}^{\rm sg} &= 0,\nonumber\\
h_{ij}^{\rm sg} & = 0,\nonumber
\end{align} 
where $\Phi_{\rm sg}$  satisfies the Poisson equation, $\nabla^{2}\Phi_{\rm sg} =4 \uppi G \rho$.

In order for this approximation to be valid, two requirements must be met:  $|h_{\mu\nu}^{\rm sg}|\ll 1$ and $|\tilde {g}_{\mu\nu} - \eta_{\mu\nu}| \ll 1$ for all elements; here $\eta_{\mu\nu}$ is the Minkowski metric.  The first condition is easily satisfied  because $|\Phi_{\rm sg}c^{-2}|\simeq (GM_{\star}/R_{\star})c^{-2}\simeq10^{-6}$.

Further steps must be taken to satisfy the second requirement because the departures from the Minkowski metric in the box frame are $\sim O(0.1)$ when the origin of our simulation box is $\simeq 20~\rg$ from the black hole ($\rg$ is the gravitational radius of the BH). Adjusting the $g_{\rm tt}$ metric element by adding the stellar self-gravity obtained from solving the Poisson equation in the box frame is then a suspect procedure.
For the purpose of combining stellar self-gravity with the global spacetime, we therefore create a new frame, one defined by an orthonormal tetrad formalism. 
The metric  in the tetrad system is, by construction, exactly Minkowski at the origin.  Elsewhere in the box, the departure of the background metric from Minkowski increases as the separation to the BH decreases. However, these departures are small in all our simulations.  They are $\sim 10^{-4}$ for $r_{\rm p}/r_{\rm g}\simeq 100$, and rise only to $\sim 10^{-3}$ at $r_{\rm p}/r_{\rm g} \approx 20$, the smallest radius reached in our simulations with $M_{\rm BH} = 10^6$.  Even at $r_{\rm p}/r_{\rm g}\simeq 5$, the smallest distance in the high black hole mass simulations of \citetalias{Ryu4+2019}, the departure from Minkowski in the tetrad frame is only $\sim 10^{-2}$.  Quantitative limits for the applicability of this approximation are presented in Appendix~\ref{appendix}.

We construct the tetrad system at the star's starting location in the usual way. We choose the time-like unit vector $\mathbf{e}_{(0)}^{\mu}$ to be the 4-velocity $\mathbf{u}^{\mu}$.
In the box frame, $\mathbf{e}_{(0)}^{\mu}=(1/\sqrt{-\tilde{g}_{00}},0,0,0)$. The remaining components $\mathbf{e}_{(i)}^{\mu}$ are found by a Gram-Schmidt method.
This procedure could be performed at each point along the orbit.  We find it more efficient, however, to perform it only once, at the starting point of the star.  Once that first system has been calculated,
we parallel-transport the tetrad basis along the star's geodesic by integrating the equation
\begin{align}
\frac{de_{(a)}^{\mu}}{d\tau} + \Gamma_{\alpha\beta}^{\mu}e_{(a)}^{\alpha}e_{(0)}^{\beta}=0, \label{eq:paralleltransport}
\end{align} 
where $e_{(0)}^{\beta}$ is the 4-velocity of the box origin and $\Gamma_{\alpha\beta}^{\mu}$ refers to the metric's affine connection evaluated in the box frame.

Both $\tilde g_{\mu\nu}$ and the tetrad basis are functions of the orbital variables $\mathbf{X}(t)$ (the star's center-of-mass position in the black hole frame) and $d\mathbf{X}(t)/dt$ (the star's coordinate velocity in the black hole frame).  Because the orbit is independent of fluid updates, we integrate the orbit of the star and the parallel-transport equation beforehand using a fourth-order Runge-Kutta integrator with adaptive time steps and make a lookup table with the orbital variables. At each time step, the code finds $\mathbf{X}(t)$ and $d\mathbf{X}(t)/dt$ from the lookup table by linearly interpolating between the two sets of data at the two most adjacent times, and then calculates $\tilde g_{\mu\nu}$. We ensure that time differences between lines of the lookup table are sufficiently small compared to the time steps for fluid updates.

The self-gravitational potential $\Phi_{\rm sg}$ of the star is computed at each step of the fluid simulation using a discrete sine Fourier transform method.  
Following \citet{ChengEvans2013}, we introduce an image mass on the box boundary so that $\Phi_{\rm sg}$ asymptotes to zero at infinity, not on the domain boundary.  Its magnitude depends on the multipole moments of the mass inside the box; we carry out the sum up to $l_{\rm max} = 4$.   We stress that this image mass is used only when calculating $\Phi_{\rm sg}$, and not when updating the fluid elements.  

Once $\Phi_{\rm sg}$ has been calculated, we add it to $\tilde{g}_{00}$ in the tetrad frame as in Equation~\ref{eq:metric}.  We then perform the inverse of the original tetrad coordinate transformation in order to find $g_{\mu\nu}$---now including both the star's and the black hole's gravity---in the box frame.  This form of the metric governs the fluid simulation.  The value of this procedure is demonstrated by contrasting the connection coefficients computed by our tetrad method with the ones found by simply solving the Poisson equation in the box frame and adding $\Phi_{\rm sg}$ to $g_{\rm tt}$: the latter method introduces errors $\approx 20 - 30 \%$ at, for example, $14r_{\rm g}$ from the black hole \footnote{ These results may be relevant to the method of \citet{Tejeda+2017}, in which stellar self-gravity is calculated by solving the Poisson equation in spherical coordinates assuming a Minkowski metric, and then adding the potential to $g_{tt}$ for Kerr spacetime.}.

\begin{figure*}
	\centering
	\includegraphics[width=6.9cm]{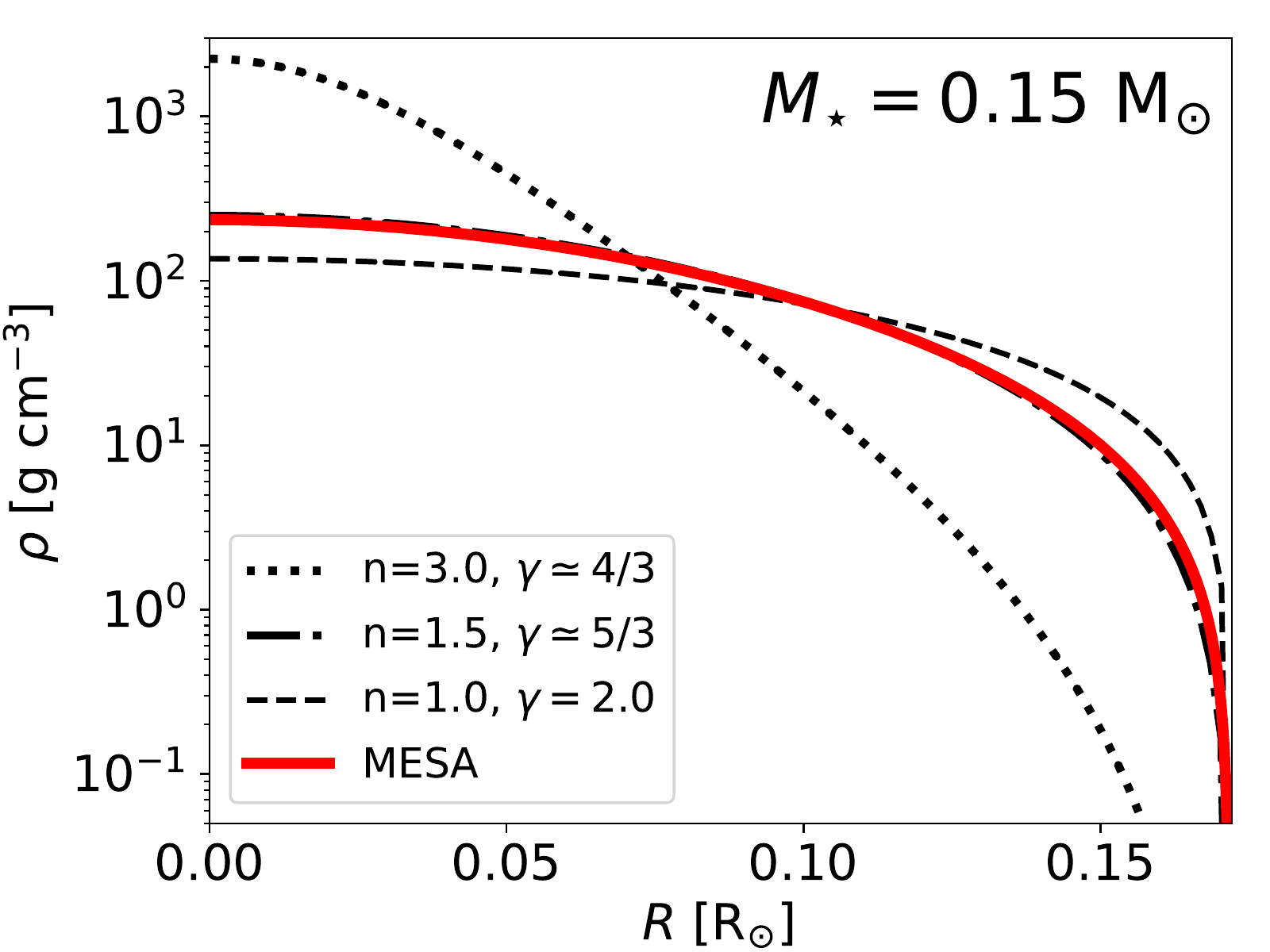}	
	\includegraphics[width=6.9cm]{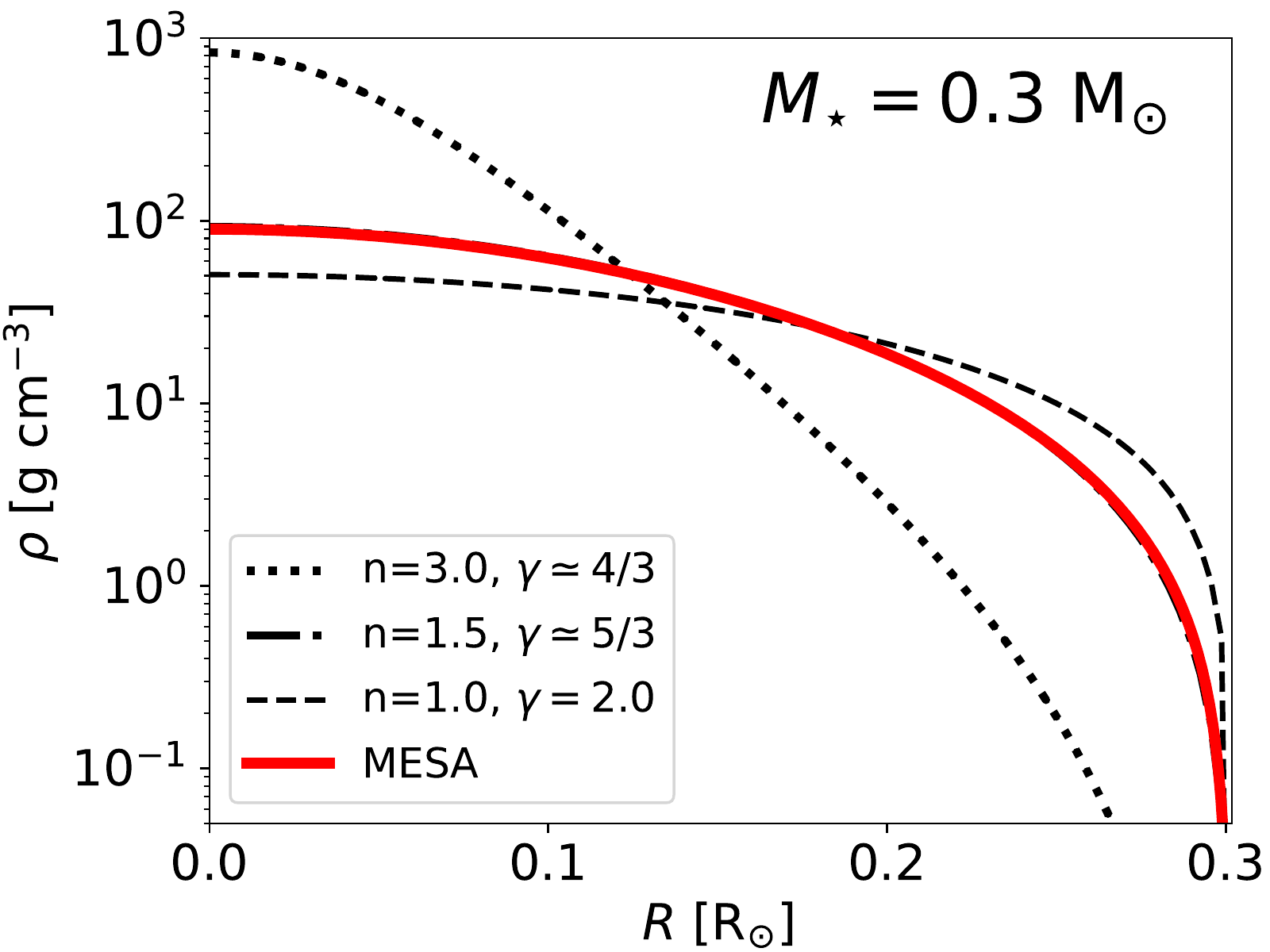}\\
	\includegraphics[width=6.9cm]{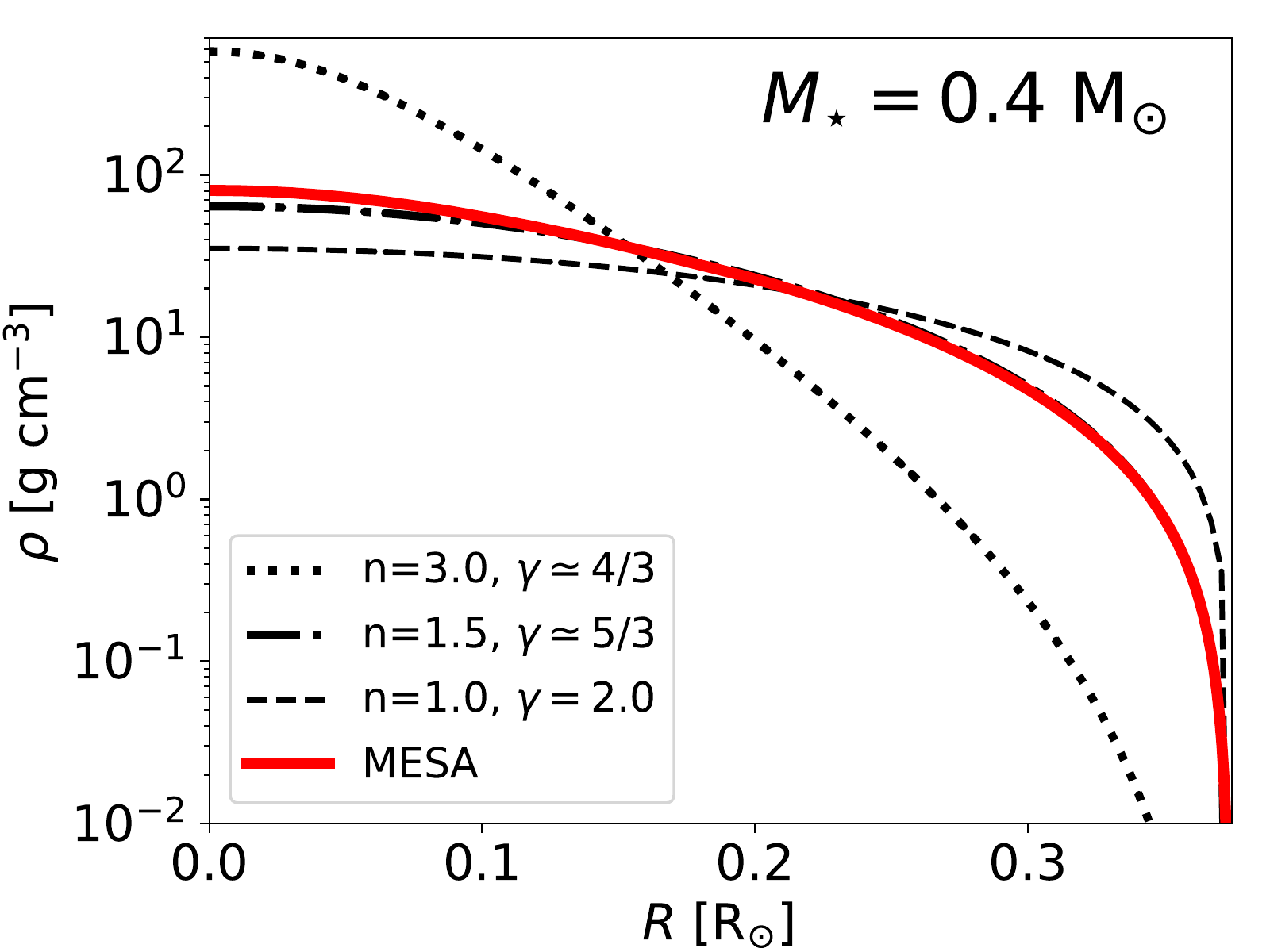}	
	\includegraphics[width=6.9cm]{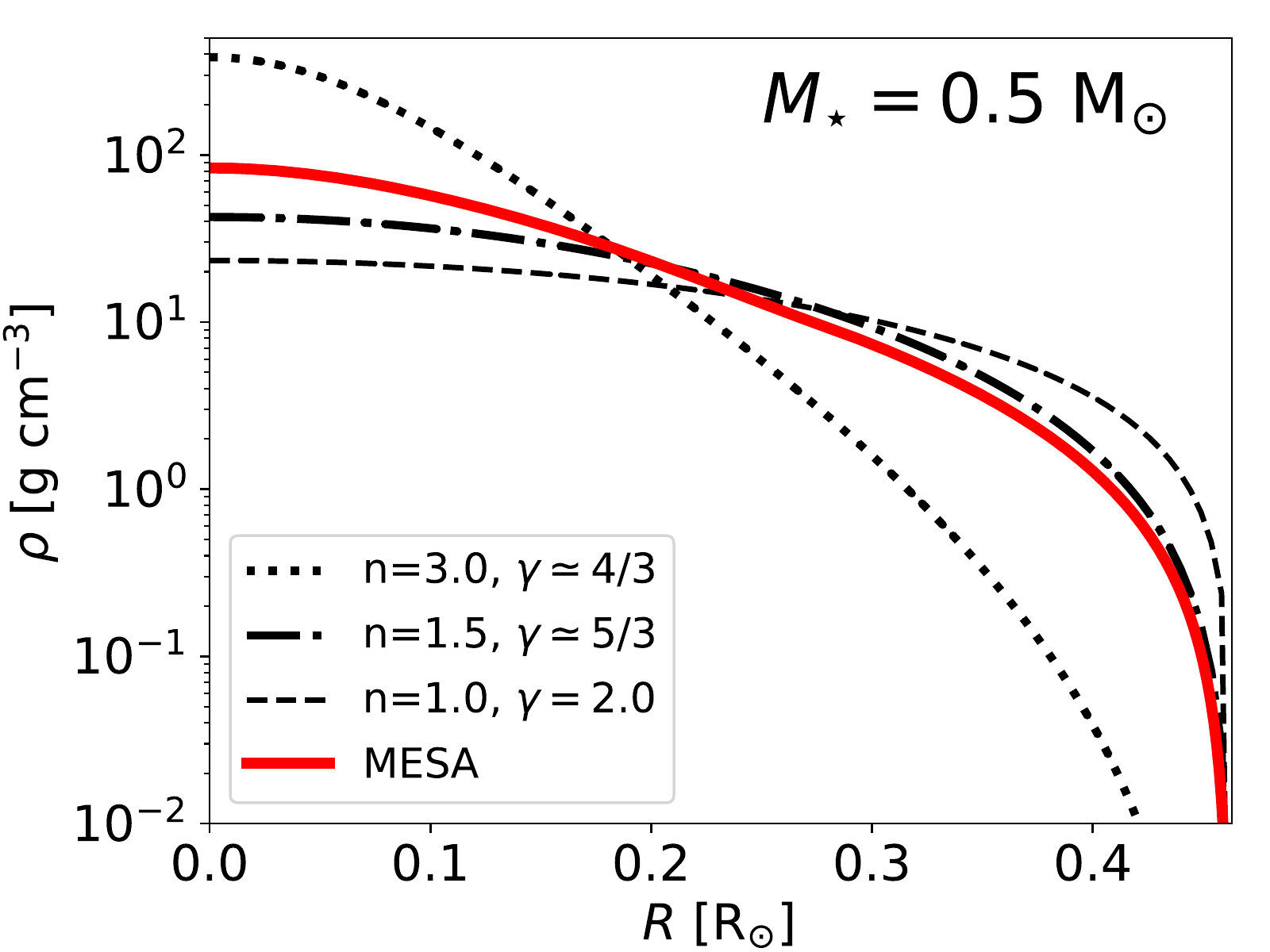}\\
		\includegraphics[width=6.9cm]{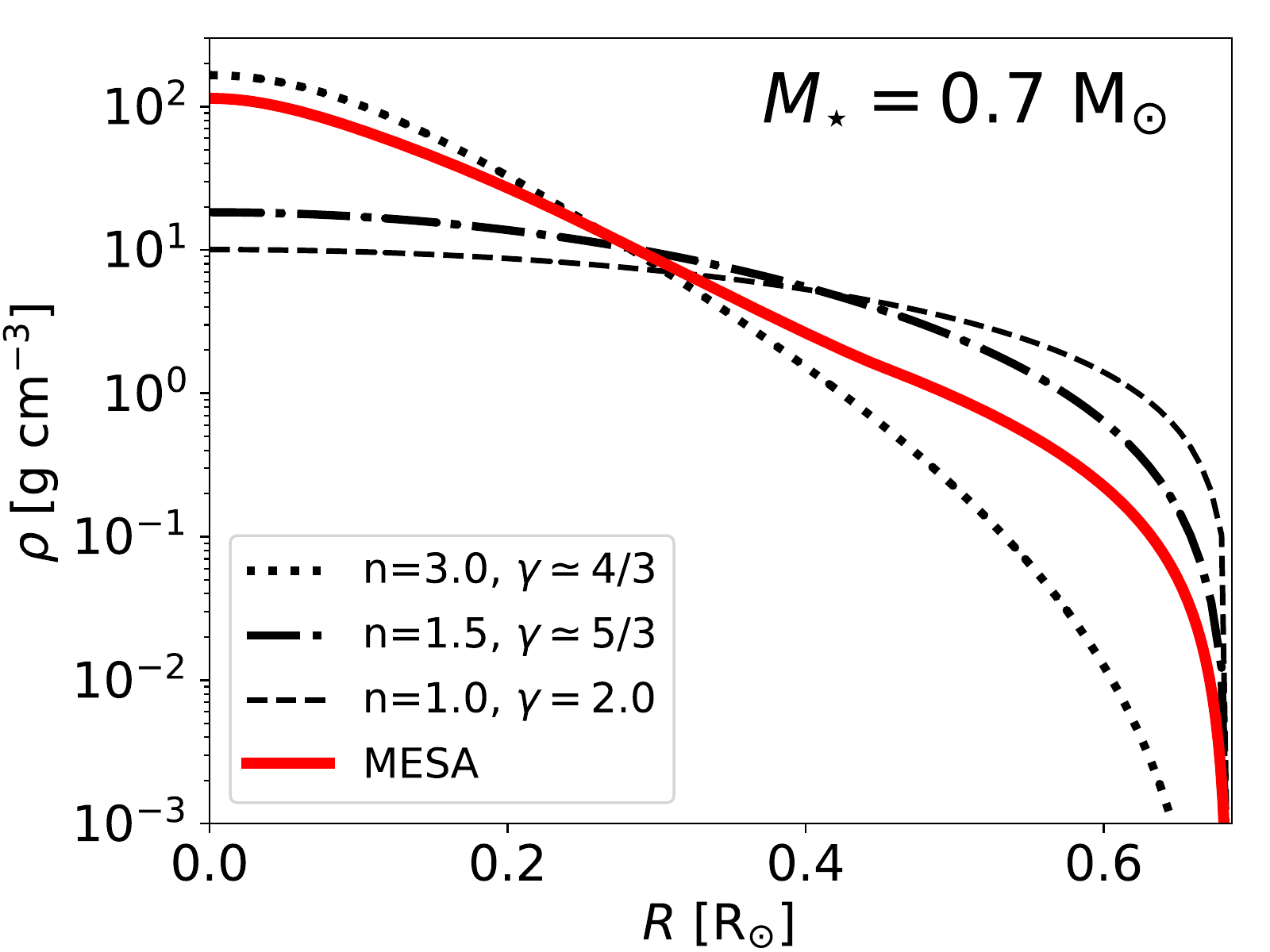}
	\includegraphics[width=6.9cm]{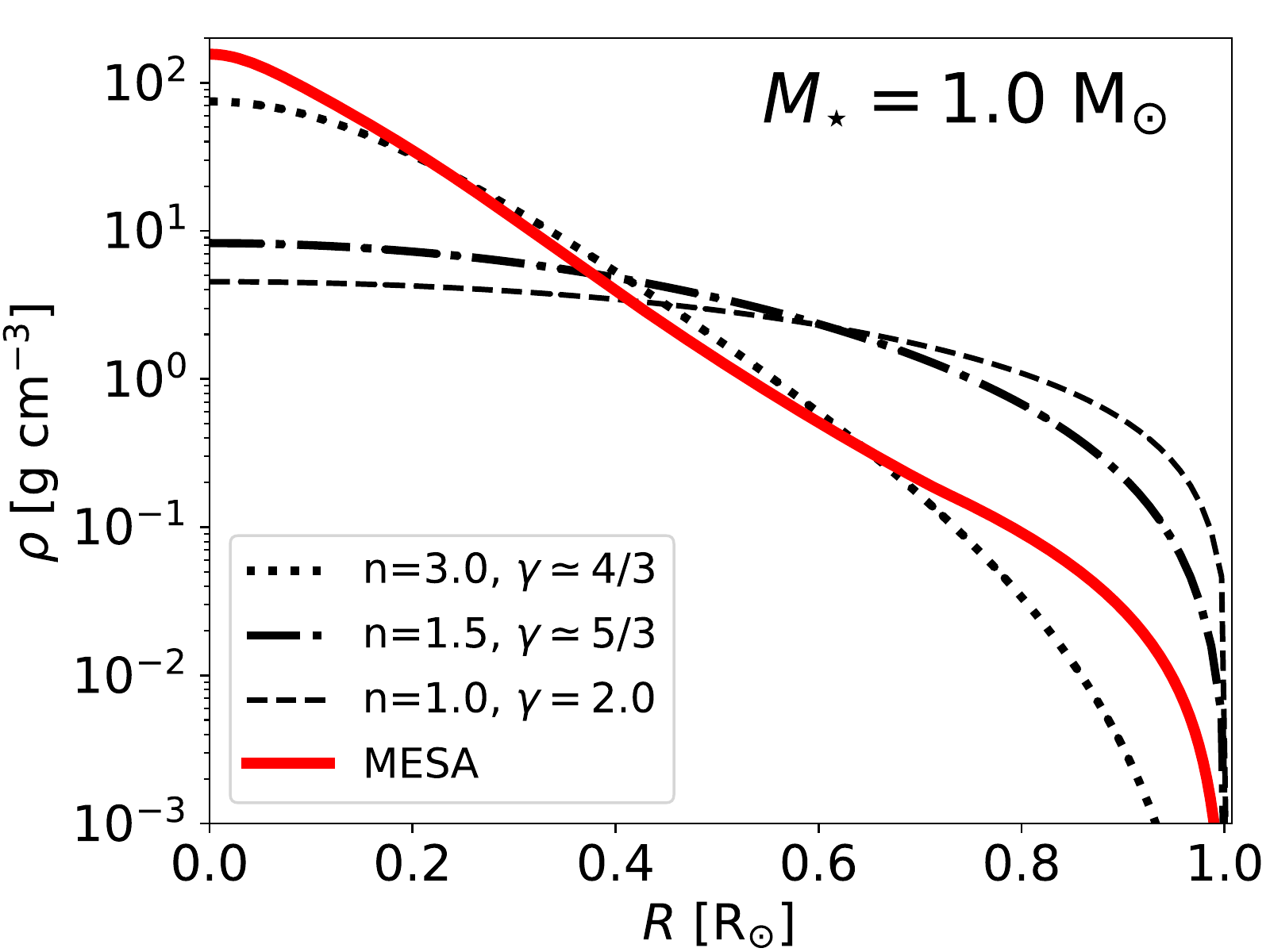}\\
	\includegraphics[width=6.9cm]{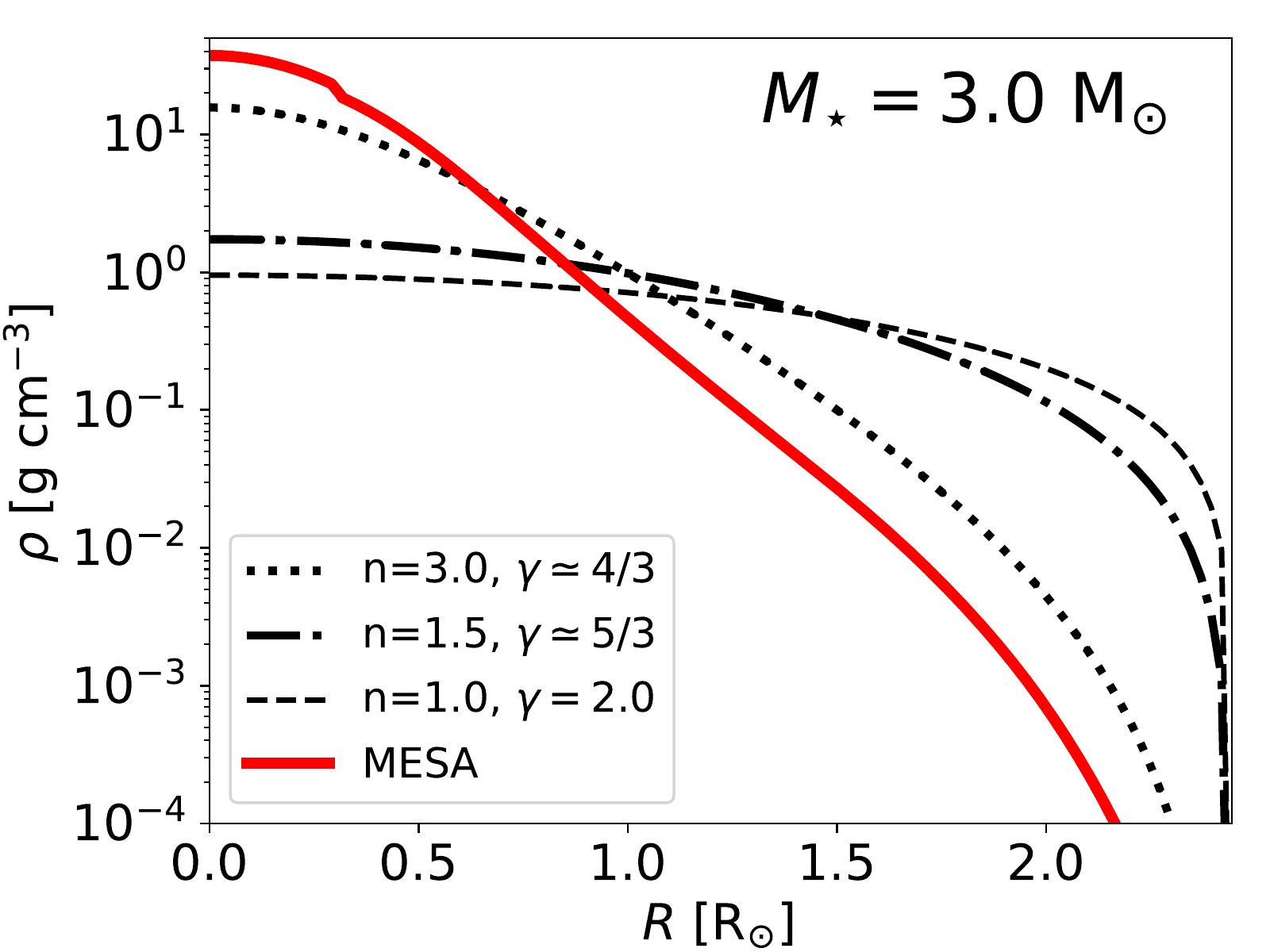}
	\includegraphics[width=6.9cm]{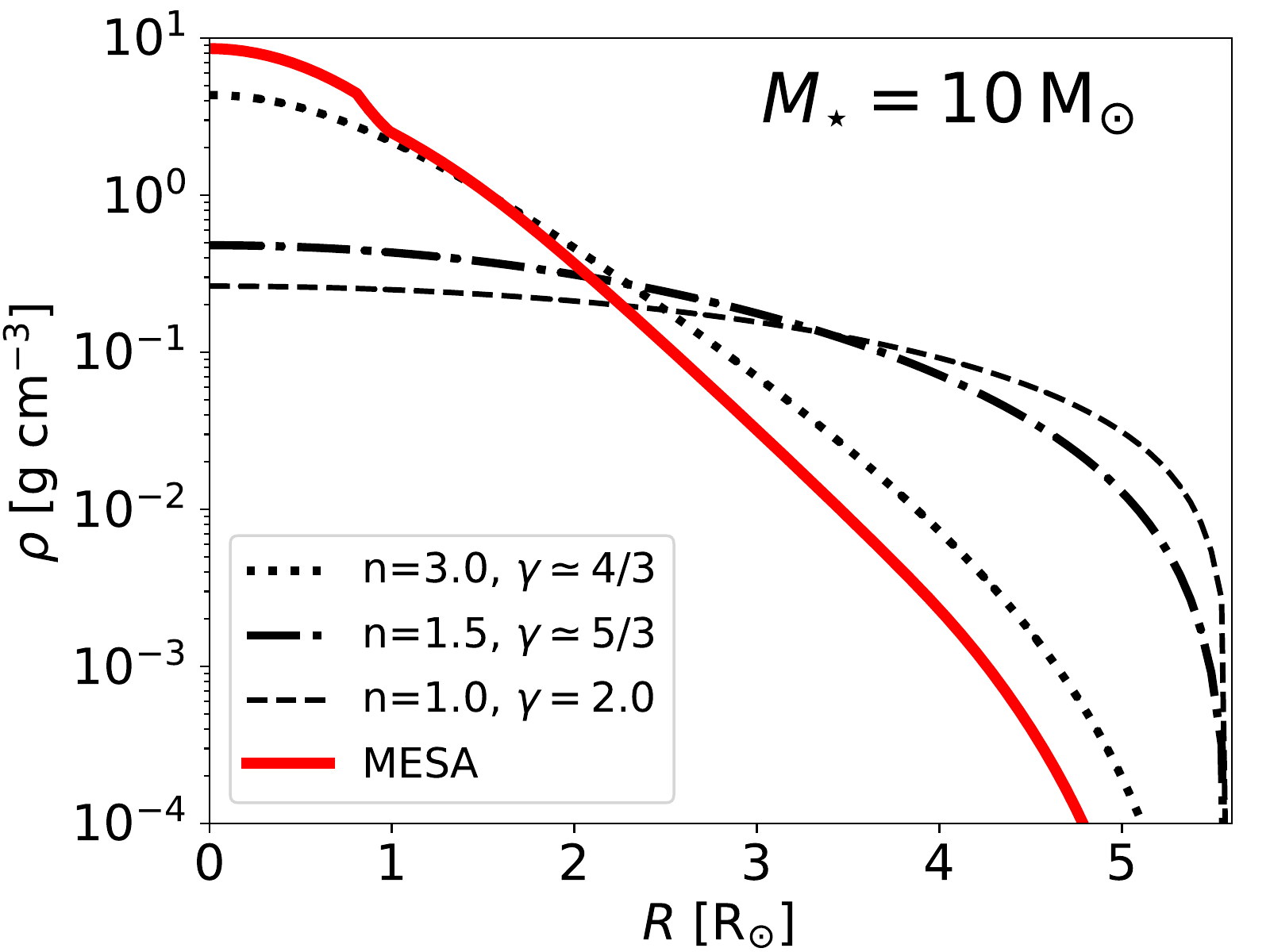}
	\caption{The radial density profiles of our MS \mesa~ models. The thick red solid lines indicate the profiles from the \mesa~ data. In each case, we show the profile only out to the radius at which we supersede the \mesa~ data in order to create a smoother connection to the external atmosphere.  The plots for the $3\Msol$ and $10\Msol$ stars have their own density scales in order to show the large range of density found in these stars. For a comparison, we overplot for each mass the density profiles predicted by polytropic models  with $\gamma= 4/3$ (dotted), $5/3$ (dot-dashed) and $2.0$ (dashed). }
	\label{fig:densityprofile}
\end{figure*}

\subsection{The stellar model}
\label{sec:stellarmodel}

To provide the initial data for our simulations, we evolve stars using the stellar evolution code \mesa~ \citep{Paxton+2011}, assuming solar metallicity, until they reach half the  MS life time for their mass. Since the life times of stars with $M_{\star}<1$ are longer than a Hubble time, we  assume all low-mass stars have an age $\sim13-14\Gyr$. 

For our suite of simulations, we consider eight MS stellar models, with masses, $M_{\star}=0.15$, $0.3$, $0.4$, $0.5$, $0.7$, $1.0$, $3.0$ and $10$. The models represent a range of different interior structures: fully convective stars ($0.15-0.3\Msol$), stars with a shallow convective envelope and a large radiative inner region ($0.4-0.7\Msol$), fully-radiative stars ($1\Msol$), and stars with a radiative envelope  but a convective core ($3\Msol$ and $10\Msol$) \citep{KippenhahnWeigert1994}. Throughout this paper we will use the term ``low-mass"  for all stars with $M_{\star}\leq0.5$, and ``high-mass" for stars with $M_{\star}\geq1$.  Within both the low-mass and high-mass groups, the internal stellar structures are similar to one another; the $M_{\star}=0.7$ structure has an intermediate character. The density profiles of these stellar models are shown in Figure \ref{fig:densityprofile}, together with a few polytropic stellar models. The $M_{\star} = 0.15$, $0.3$ and $0.4$ stellar models are in good agreement with a polytropic model with $\gamma= 5/3$ for the given mass and radius. The $M_{\star} =1$ star is closely matched by a polytrope with $\gamma=4/3$ at intermediate radii, but not near the core or the surface. The other stars do not resemble any polytropic model. Stars with $M_{\star}\geq1$ tend to have a more concentrated inner region than low-mass stars or polytropic stars with $\gamma=4/3$. 
We summarize the model parameters of the MS stars in Table \ref{tab:modelparameter}.

We find that the relation between $M_{\star}$ and $R_{\star}$ for $0.15\leq M_{\star}\leq 3$ is well-described by the formula\footnote{ This approximate formula is a best fit over the whole range, hence it is not normalized at $M_{\star}=1$.}
\begin{align}\label{eq:M_R_relation}
R_{\star}&= 0.93~M_{\star}^{0.88}.
\end{align}
The fractional differences between $R_{\star}$ estimated using Equation~\ref{eq:M_R_relation} and  $R_{\star}$ taken from the MESA models are all less than $0.1$ for $M_{\star} \leq 3$, but for $M_{\star}=10$, the fractional difference is 0.27. This relation is consistent with that of \citet{KippenhahnWeigert1994} even though they found $d\ln R_{\star}/d\ln M_{\star} \simeq 0.8$ for low-mass stars and $d\ln R_{\star}/d\ln M_{\star} \simeq 0.6$ because those slopes did {\it not} apply to $M_{\star} \sim 1$, where the slope was rather higher.

\subsection{Running the Simulations}\label{sec:procedure}

\subsubsection{Initial stellar structure}
\label{sec:initial}

At the start of each simulation, a MS star in hydrostatic equilibrium is placed so that its center lies at the coordinate origin of the box. The density and pressure profiles of the star are determined by a linear interpolation between two adjacent data points in the \mesa~model whose positions are closest to each cell center of our \harm~grid. After doing so, the profiles on our grid agree with the  \mesa~profiles to within less than $0.1\%$ out to a radius at which the enclosed mass $\simeq99\%~M_{\star}$.

To avoid creating too sharp a discontinuity between the stellar density and the external ``vacuum", we extrapolate the logarithmic density gradient at the $99\%$ mass radius to larger radii,
 but not permitting the density to fall below the vacuum density. To ensure that the extrapolation does not affect our results, as mentioned in Section~\ref{subsec:computationaldomain}, we choose the initial distance of the stars from the BH to be sufficiently great that the stellar configuration is completely relaxed long before the star approaches pericenter. The  pressure in the extrapolation region is determined by the hydrostatic equilibrium condition with a temperature comparable to the stellar surface temperature.
We set the vacuum density low enough to ensure that the total mass of the domain, minus $M_{\star}$, is $< 10^{-3} M_{\star}$. The simulation's absolute density floor is $(10^{-1}-10^{-2})\times$ the vacuum density. 

 Once the evolution begins, any small departures from hydrostatic balance in the outer 1\% of mass are relaxed away on the vibrational timescale $\tau_{\star}$, which we define as $\tau_{\star} = \left(3GM_{\star} / 4\pi R_{\star}^3\right)^{-1/2}$. 
Kept far from the black hole, these stellar models stay in hydrostatic equilibrium for much longer than the time it takes for the stars to pass the pericenter, i.e., $>25~\tau_{\star}$.

\subsubsection{Stellar trajectories}
\label{sec:stellartrajectories}

For each stellar model, we select a number of parabolic Schwarzschild geodesics  with different pericenter distances in order to explore the transition from partial to full disruption. For those reported in this paper, all have $M_{\rm BH}=10^6$.  
We label them by the penetration factor $\beta \equiv r_{\rm t}/r_{\rm p}$.
We provide the value of $\beta$ in Table~\ref{tab:modelparameter}.

For stars with $M_{\star}<0.7$, we consider $\beta$ in the range $0.5<\beta<1.2$; for higher-mass stars,  $0.67<\beta<2.86$), with a small shift toward larger $\beta$ for stars with larger mass (see  Table~\ref{tab:modelparameter}).  In every case, the initial distance of the star from the BH is $\simeq10\rtidal$; with this choice, the star passes through pericenter at $t\simeq8~\tau_{\star}$. 
We continue to follow the event until the center-of-mass of the star reaches $r\simeq 20-30~r_{\rm t} $.  At this point in all our runs, it has become clear whether the event results in a total disruption or a partial one, and if partial, the mass of the remnant is well-determined.

\begin{table}

								\centering
	\renewcommand{\thetable}{\arabic{table}}
	\caption{Model parameters of MS stars considered in this study. The MS stars are evolved using \mesa~ until their ages become half the typical MS life times. Their vibration time $\tau_{\star}$ is defined as $\tau_{\star}=1.0/\sqrt{GM_{\star}/(4\uppi R_{\star}^3/3)}$. We list the order-of-magnitude tidal radius $r_{\rm t}$ of each star, $\beta\equiv r_{\rm t}/r_{\rm p}$ considered in our TDE experiments.} \label{tab:modelparameter}
	\begin{tabular}{c c c c c}
		\tablewidth{0pt}
		\hline\noalign{\smallskip}
		\hline\noalign{\smallskip}
		$M_{\star}^{a}$ &  $R_{\star}^{b}$ & $\tau_{\star}^{c}$  & $\rtidal/r_{\rm g}$& $\beta=r_{\rm t}/r_{\rm p}$ \\
		\hline\noalign{\smallskip}
		$0.15$   & $0.17$   &  $0.6$   & $15$  & 0.50, 0.71, 0.67, 0.63, 0.56, 1.00 \\\noalign{\smallskip}	
		$0.30$   & $0.30$   &  $1.0$   & $21$  & 1.0, 0.83, 0.77, 0.71, 0.67, 0.56 \\\noalign{\smallskip}
		$0.40$   & $0.37$    &  $1.2$   & $24$ & 1.0, 0.83, 0.77, 0.71, 0.67, 0.56 \\\noalign{\smallskip}	
		$0.50$   & $0.46$    &  $1.5$   & $27$ & 1.25, 1.00, 0.91, 0.83, 0.67, 0.56 \\\noalign{\smallskip}
		$0.70$   & $0.69$    &  $2.2$   & $36$ & 1.67, 1.54, 1.43, 1.25, 1.11, 0.67 \\\noalign{\smallskip}	
		$1.0$   & $1.0$     & $3.3$  & $47$ & 2.50, 2.22, 2.00, 1.82, 1.49, 1.79 \\\noalign{\smallskip}
		$3.0$   & $2.4$    &  $7.2$ & $80$  & 2.86, 2.50, 2.22, 2.00, 1.67, 1.18 \\\noalign{\smallskip}
		$10$    &  $5.6$     & $14$ & $120$ & 2.86, 2.50, 2.22, 2.00, 1.67, 1.18 \\\noalign{\smallskip}
		\hline\noalign{\smallskip}
		\multicolumn{5}{l}{\small Units : $^{a}$ $\mathrm{M}_{\odot}$; $^{b}$ $\mathrm{R}_{\odot}$; $^{c}$ $10^{3}\s$.}
	\end{tabular}
\end{table}

\begin{table}
	\renewcommand{\thetable}{\arabic{table}}
	\centering
	\caption{The physical tidal radii $\mathcal{R}_{\rm t}$ for MS stars encountering a $10^{6}\Msol$ non-spinning black hole. 
	The errors related to $\physrad$ originate from the finite sampling of pericenter as shown in Table~\ref{tab:modelparameter}. In the last column $\Delta E/\Delta\epsilon$ is the ratio of the actual characteristic debris energy width (containing 90\% of the total mass) to the order-of-magnitude estimate.} \label{tab:tidal_r}
	\begin{tabular}{c c c c c c}
		\tablewidth{0pt}
		\hline
		\hline \noalign{\smallskip}
		$M_{\star}^{a}$ &  $R_{\star}^{b}$  & $r_{\rm t}/\rg$
		& $\mathcal{R}_{\rm t}/\rg$ & $\Psi=\mathcal{R}_{\rm t}/r_{\rm t}$
		& $\Delta E/\Delta\epsilon$ \\ \noalign{\smallskip}
		\hline \noalign{\smallskip}
		$0.15$  & $0.17$    & $15.2$&         $22.1\pm0.8$    &  $1.45\pm0.05$ 
		&    0.67        \\\noalign{\smallskip}
		$0.30$   & $0.30$& $21.2$  &  $26.5\pm1.1$   & $1.25\pm0.05$ 
		& 0.75\\\noalign{\smallskip}
		$0.40$   & $0.37$ & $23.9$ & $30.0\pm1.4$   & $1.25\pm0.05$ 
		& 0.75\\		\noalign{\smallskip}
		$0.50$   & $0.46$ & $27.4$   &$28.9\pm1.4$   & $1.05\pm0.05$ 
		& 0.85 \\\noalign{\smallskip}
		$0.70$   & $0.69$ & $36.4$  &$24.6\pm1.4$   & $0.675\pm0.025$ 
		&1.09 \\		\noalign{\smallskip}
		$1.0$   & $1.0$       & $47.5$ &$22.5\pm1.2$  & $0.475\pm0.025$& 1.47\\\noalign{\smallskip}
		$3.0$   & $2.4$      & $79.8$ & $33.9\pm2.0$ & $0.425\pm0.025$
		& 1.83\\\noalign{\smallskip}
		$10$    &  $5.6$      & $123$  & $52.1\pm3.1$ & $0.425\pm0.025$
		& 1.79\\\noalign{\smallskip}
		\hline
				\multicolumn{5}{l}{\small Units : $^{a}$ $\mathrm{M}_{\odot}$; $^{b}$ $\mathrm{R}_{\odot}$.}
	\end{tabular}
\end{table}

\subsubsection{Distinguishing partial from full disruptions and determination of the physical tidal radius}\label{subsec:disruptioncondition}

We define complete disruption of a star as the satisfaction of three criteria at the end of a simulation.  Without exception, decisions made on the basis of these criteria are consistent. 

\begin{figure}
	\centering
	\includegraphics[width=8.6cm]{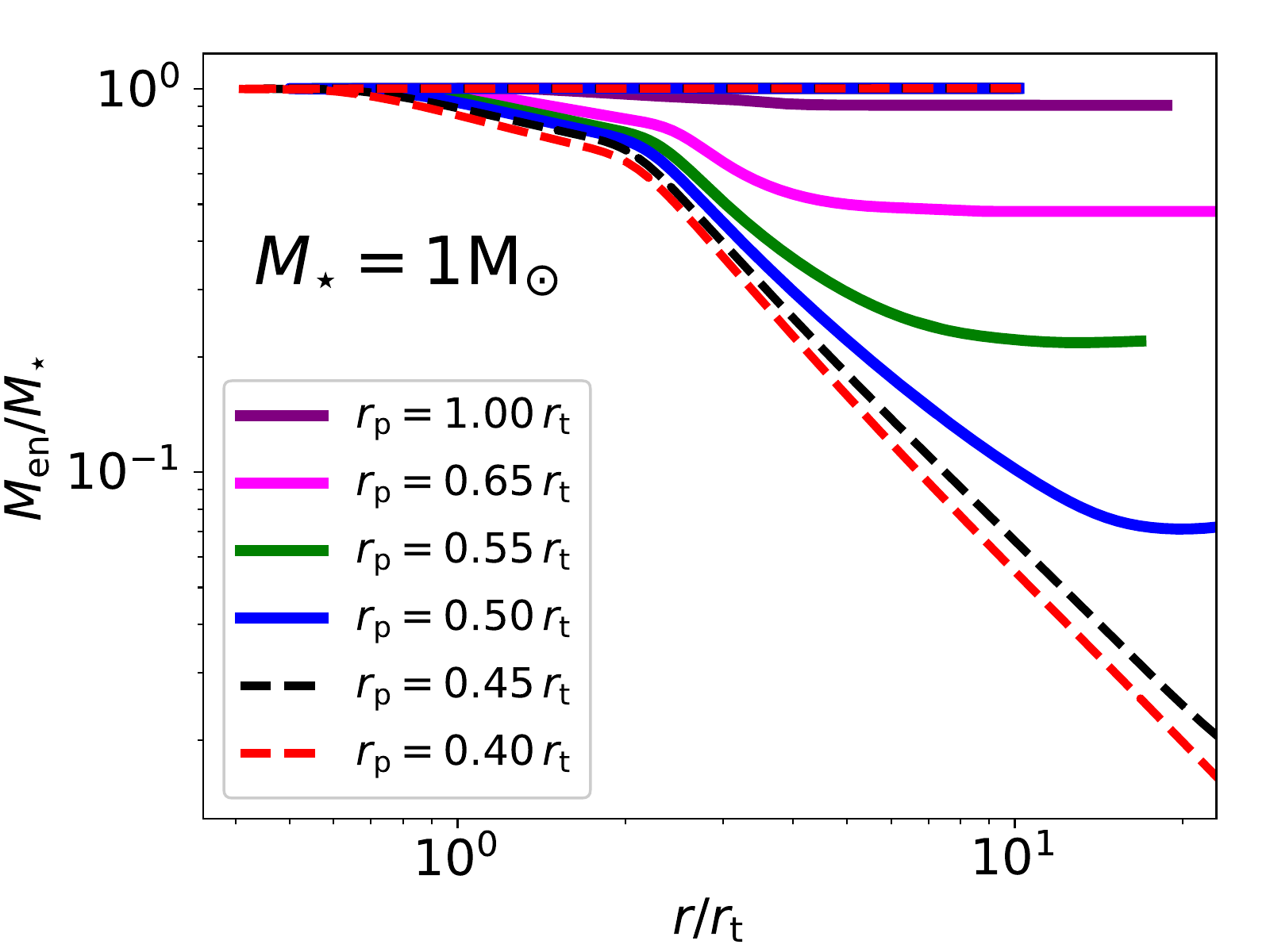}		
	\caption{The fraction of initial stellar mass $M_{\rm en}/M_{\star}$ enclosed in the box versus the box's radial position over time, $r(t)$, for a $1\Msol$ star undergoing  tidal encounters at various periastron distances. Continuous decrease in mass for complete disruptions (dashed lines) is clearly distinguished from the mass change for partial disruptions (solid lines). Notice that the perfectly flat lines at $M_{\rm rem}/M_{\star}=1$ show that the stars are in hydrostatic equilibrium before tidal forces affect the stars.}
	\label{fig:massfraction}
\end{figure}

\begin{enumerate}
	\item \label{con1}  Lack of any approximately spherical bound structure.

	\item \label{con2} Monotonic (as a function of time) decrease in the maximum pressure of the stellar debris. 
	
	\item \label{con3} Monotonic decrease in the mass within the computational box. This criterion is illustrated in Figure \ref{fig:massfraction}.   The mass remaining in the box for complete disruption falls with increasing distance from the BH $\propto r^{-\alpha}$ with $\alpha\simeq1.5-2.0$, whereas for partial disruptions the remaining mass eventually becomes constant, which signifies a persistent self-gravitating object. 
	
\end{enumerate}
Once all encounters for a given $M_{\star}$ and $M_{\rm BH}$ are identified as either full or partial, we estimate the physical tidal radius $\physrad$ as the mean of the largest $r_{\rm p}$ yielding a full disruption and the smallest $r_{\rm p}$ producing a partial disruption.  Consequently, the uncertainty of $\physrad$ originates from the discrete sampling of $\beta$.

\section{Results}
\label{sec:results}

\subsection{The physical tidal radius}
\label{sec:physrtidal}

\begin{figure}
	\centering
	\includegraphics[width=8.9cm]{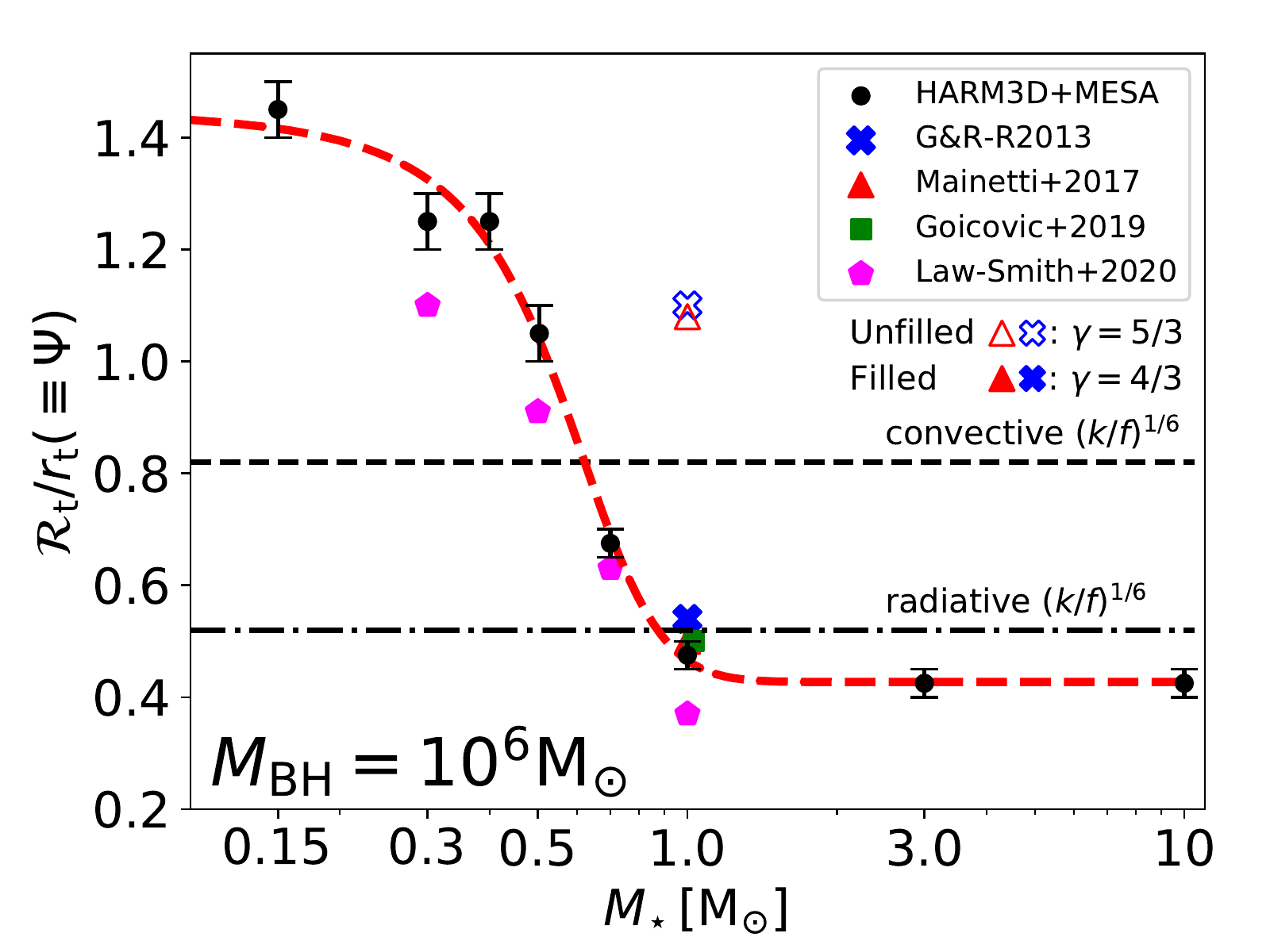}
	\caption{$\mathcal{R}_{\rm t}/r_{\rm t}(\equiv\Psi)$ for the model stars. The error bars indicate the uncertainties of $\physrad$ due to the discrete sampling of $r_{\rm t}/r_{\rm }$. The horizontal lines indicate the predictions from the ratio of apsidal motion constant $k$ and dimensionless binding energy $f$ : $(k/f)^{1/6}=0.82$ (dashed line) for low-mass stars and $0.52$ (dot-dashed line) for high-mass stars \citep{Phinney1989}. We also mark $\Psi$ found in \citet[][G\&R-R2013, blue cross]{Guillochon+2013}, \citet[][red triangle]{Mainetti+2017}, \citet[][zero-age  main  sequence, green square]{Goicovic+2019} and \citet[][middle-age main sequence, magenta pentagon]{Law-Smith+2020}. Notice that for a better distinction between dots near $\Psi\simeq0.5$ at $M_{\star}=1$, we horizontally shift the red and green dots by a small amount ($\pm0.03$). For the polytropic models (triangles and crosses), the hollow (solid) markers refer to $\Psi$ for $\gamma=5/3$ ($\gamma=4/3$).  
	The red dotted curve depicts the fitting formula (Equation~\ref{eq:r_tidal_fit_both}) introduced in \citetalias{Ryu1+2019}.}
	\label{fig:tidal_distance}
\end{figure}

The first product of our simulations is the distinction
between those pericenters yielding partial disruptions and those yielding full disruptions.  Not surprisingly, the classic tidal radius estimator $\rtidal$ is good at the order-of-magnitude level, but does not indicate the {\it physical} tidal radius (the divide between partial and full disruptions) to better than a factor of 2 (as already indicated by earlier Newtonian simulations of polytrope approximations such as \citealt{Guillochon+2013}).  What is new here is to find that the quantitative corrections are also affected by both non-polytropic internal structure (see Section~\ref{subsub:comparison_rt}) and  relativistic effects that strengthen with increasing $M_{\rm BH}$ (see \citetalias{Ryu1+2019} and \citetalias{Ryu4+2019}). As shown in Table~\ref{tab:tidal_r} and Figure~\ref{fig:tidal_distance}, the ratio $\Psi \equiv \mathcal{R}_{\rm t}/r_{\rm t}$ rises to $\simeq 1.4$ for extremely low mass ($M_{\star}=0.15$), drops gradually as the mass increases to $M_{\star}\simeq 0.5$, and then drops rapidly to $\simeq 0.4$--0.45 for $M_{\star} > 1$. Remarkably, as discussed in \citetalias{Ryu1+2019}, $\mathcal{R}_{\rm t}/r_{\rm g} \simeq 27$ for $M_{\rm BH} = 10^6$ nearly independent of $M_{\star}$ from $M_{\star} = 0.15$ to $M_{\star} \simeq 3$. As also reported in \citetalias{Ryu1+2019}, 
$\Psi$ can be expressed separately in terms of $M_{\rm BH}$- and $M_{\star}$-dependent terms.  We define the $M_{\star}$-dependent term, denoted by $\Psi(M_{\star})$, to match $\Psi$ for $M_{\rm bH}=10^{6}$.  It is well-fit by
\begin{align}
\Psi(M_{\star})& = \frac{1.47+ ~\exp[(M_{\star}-0.669 )/0.137]}{1 + 2.34~\exp[(M_{\star}-0.669)/0.137]}.
\label{eq:r_tidal_fit_both}
\end{align}
In Section~\ref{subsec:analyticmodel}, we show that $\mathcal{R}_{\rm t}$ can be estimated---without extensive simulation---by comparing the effective density of the black hole $M_{\rm BH} / \mathcal{R}_{\rm t}^3$ to the central density of the star $\rho_{\rm c}$ (Equation~\ref{eq:Rt-zeta}).

\begin{figure}
	\centering
	\includegraphics[width=8.9cm]{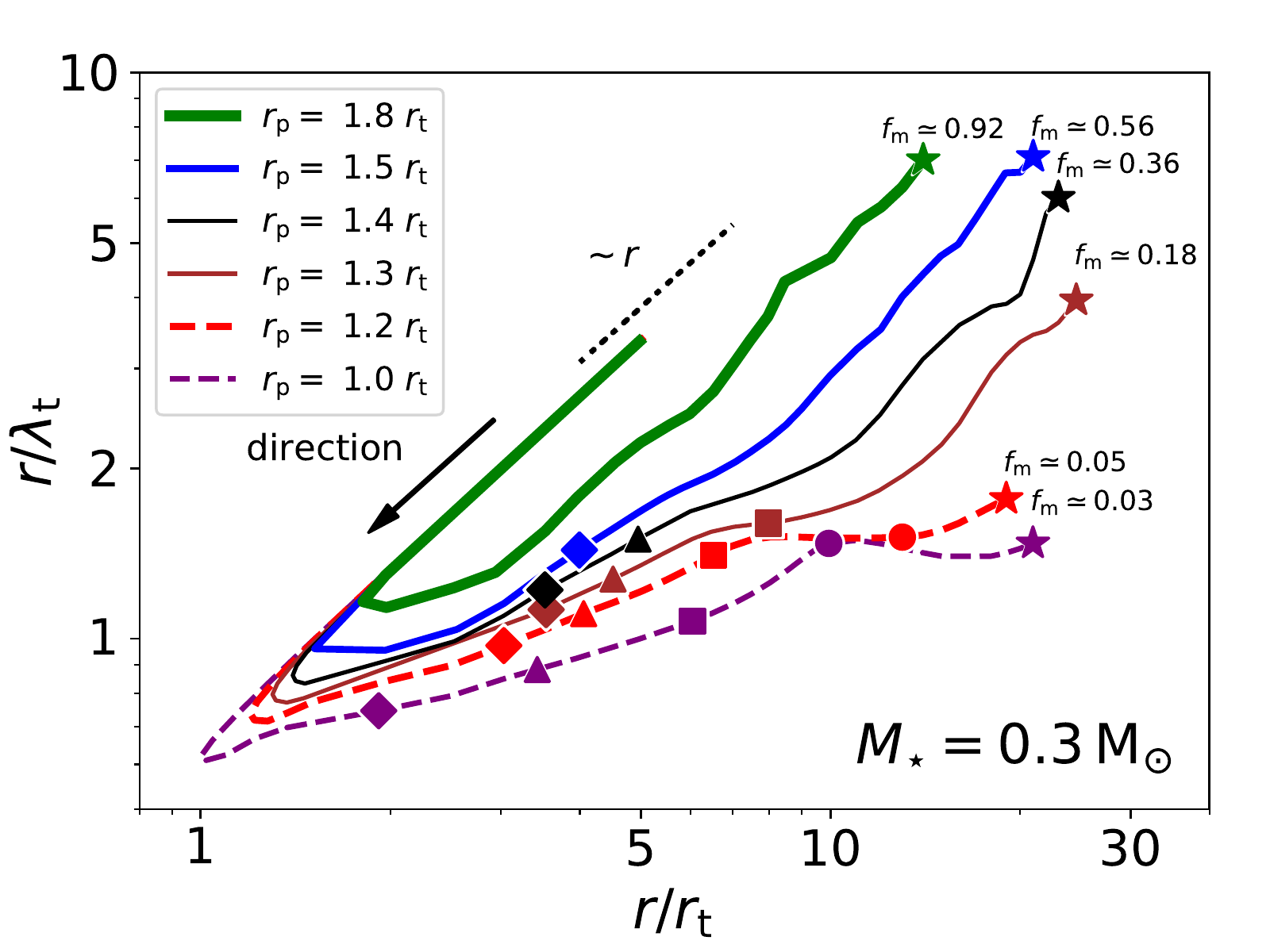}			
	\caption{The locations of fully (dashed) and partially (solid) disrupted $0.3\Msol$ stars with respect to the effective tidal sphere $\lambda_{\rm t}$ (see Equation~\ref{eq:effect_rt}). The diagonal black arrow, pointing left bottom corner, indicates the direction of motion. The markers represent the ratio of the mass retained in the box to $M_{\star}$, denoted by $f_{\rm m}$, indicating how the mass in the box depends on time: diamond ($f_{\rm m}=0.75$), triangle ($f_{\rm m}=0.5$), square ($f_{\rm m}=0.25$), circle ($f_{\rm m}=0.10$) and star ($f_{\rm m}$ at the end of simulation). The diagonal dotted line depicts the case when the average density is constant, $r/\lambda_{\rm t}\propto r$ where $r$ refers to the Schwarschild radial coordinate of the star's center of mass.}
	\label{fig:effective_rt}
\end{figure}

\subsection{Duration of tidal disruption}
\label{subsub:duration}

The classic order-of-magnitude estimate of the tidal radius amounts to the statement that at $\rtidal$ the Newtonian tidal gravity of the black hole should match the self-gravity at the surface of the star.  At the qualitative level, this comparison divides the realm of strong and weak tidal forces.  However, because stars lose mass during a tidal encounter while also changing their distance from the black hole, the sense of this comparison can be a function of time. To study how it evolves through an event, we introduce a quantity we call the ``instantaneous tidal radius" that can be measured in our data:
\begin{align}
\lambda_{\rm t}(r) \equiv \left (\frac{M_{\rm BH}}{\doverline{\rho}(r)}\right)^{1/3},
\label{eq:effect_rt}
\end{align}
where  $\doverline{\rho}$ is the average density of the cells containing $99\%$ of the total mass in the domain when summed outward from the center.

Figure~\ref{fig:effective_rt} shows how the distance of a star from the black hole in units of $\lambda_{\rm t}$ changes as a function of its distance from the black hole in units of $r_{\rm t}$.  Although the example we show is for a $0.3\Msol$ star, the same diagram for other masses is qualitatively very similar.
The lines are all initially straight because the incoming stars stay intact, i.e.,  $\doverline{\rho}$ remains constant, so that $\lambda_t$ is likewise constant, and $r/\lambda_t \propto r$. However, there is a noticeable contrast between the behavior of full and partial disruptions. When the encounter ends in the complete dissolution of the star, after the star passes pericenter, $r/\lambda_{\rm t}$ increases quite slowly, approximately $\propto r^{1/3}$, and it remains near unity out to $r\gtrsim20\rtidal$.   On the other hand, when the ultimate result is a partial disruption, after pericenter passage $r/\lambda_{\rm t}$ is also $\propto r^{1/3}$, much like the full disruption tracks, but with a larger coefficient. However, this slope ends earlier, steepening sharply when $r \gtrsim10\rtidal$ (\citealt{Steinberg+2019} find a similar result for full disruptions in which $r_{\rm p}\ll r_{\rm t}$, but the outgoing track is slightly steeper: $r/\lambda_{\rm t}\propto r^{1/2}$).  
 
The same curves also show the pace of mass-loss. Both full and partial disruptions exhibit mass-loss during the entire period when $r/\lambda_{\rm t} \sim 1$.  In partial disruptions, mass-loss continues until the star has reached $\sim10\rtidal$, while mass loss continues  until $r$ is at least $\sim20\rtidal$ in full disruptions.
In other words, mass is lost for as long as $r \sim \lambda_{\rm t}$, and this state can endure for as long as the time required for the star to swing from $r_{\rm p}$ to 10--$20\rtidal$.

\begin{figure}
	\centering
	\includegraphics[width=8.9cm]{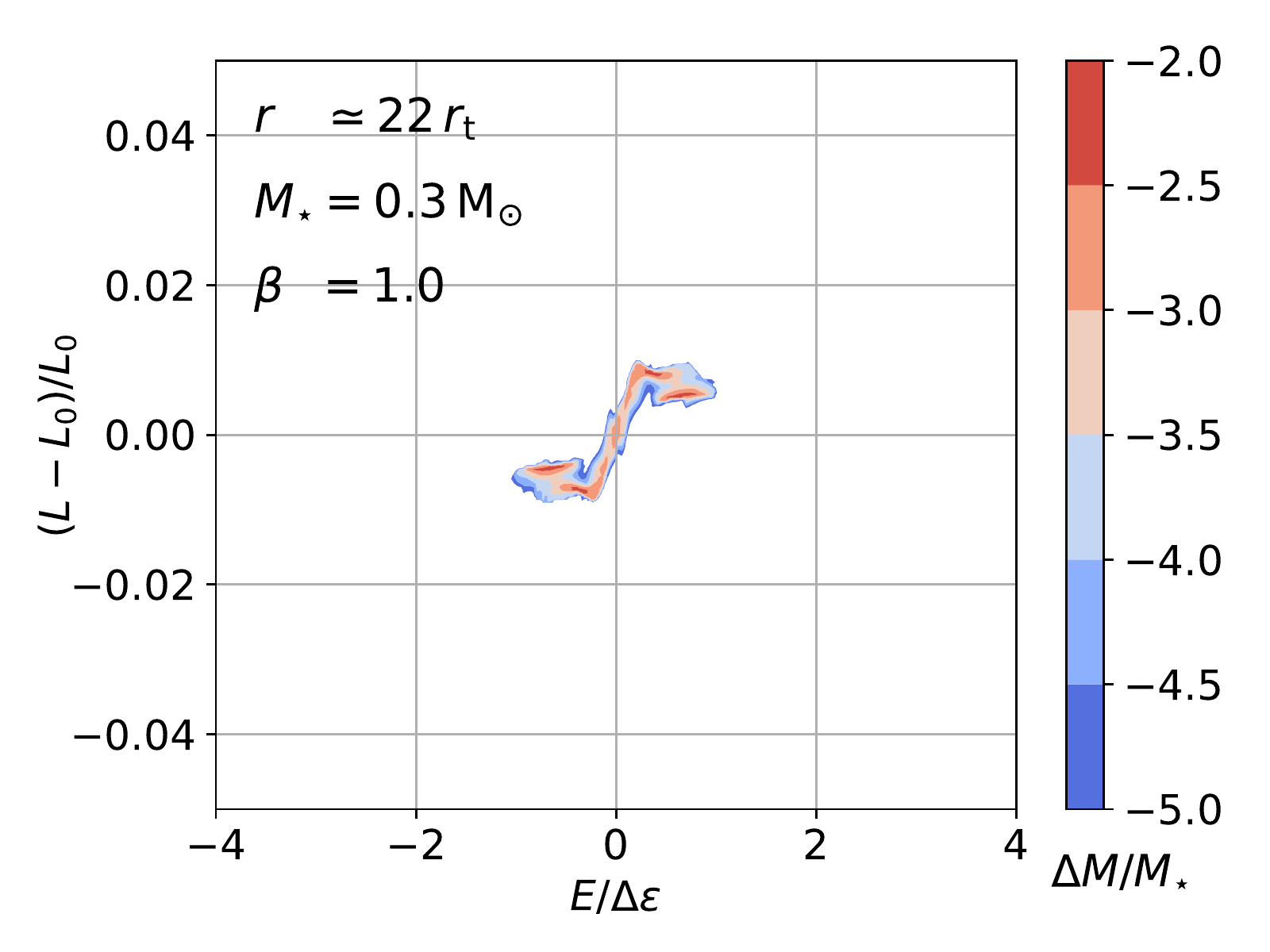}		
	\includegraphics[width=8.9cm]{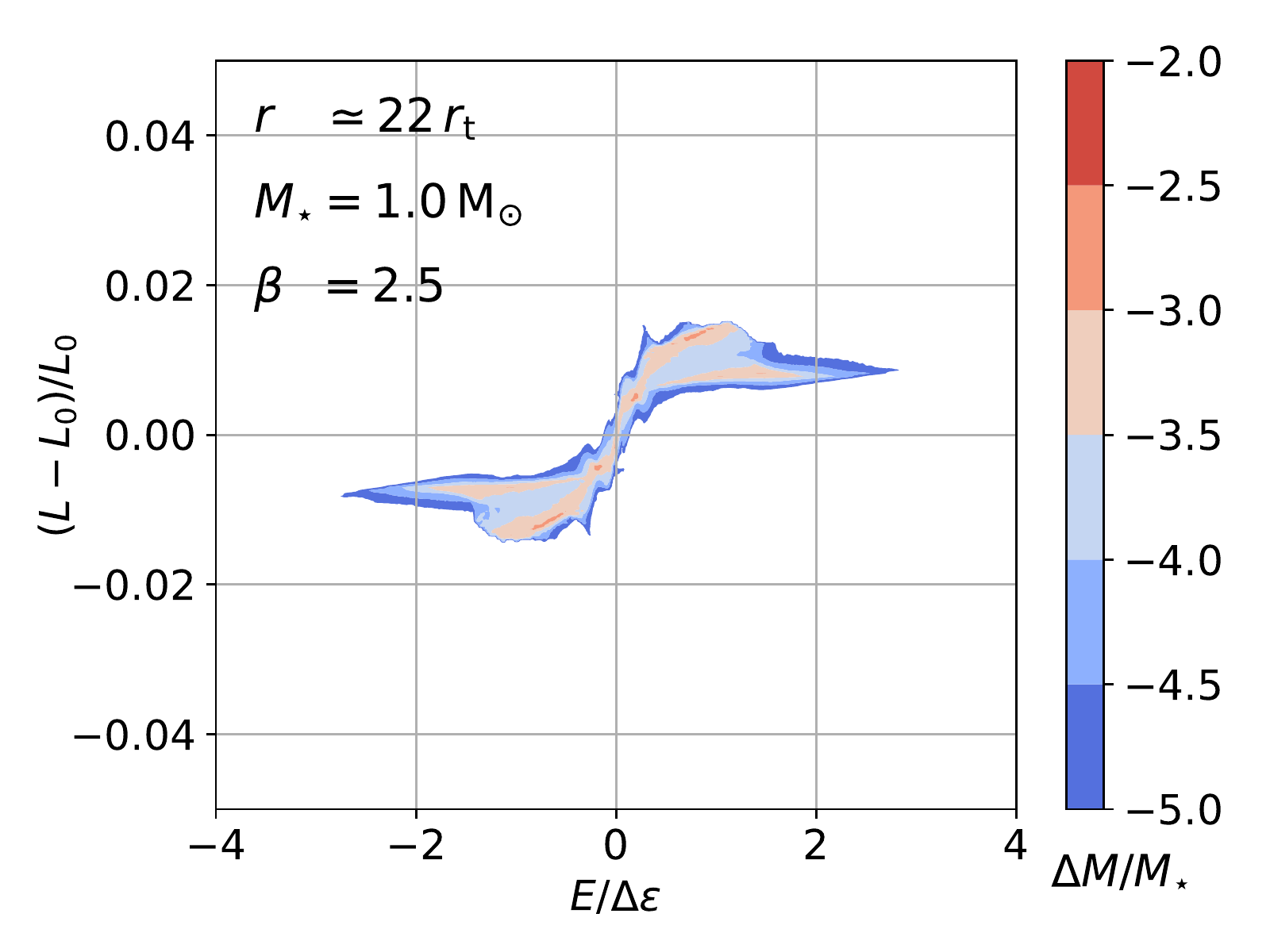}
	\includegraphics[width=8.9cm]{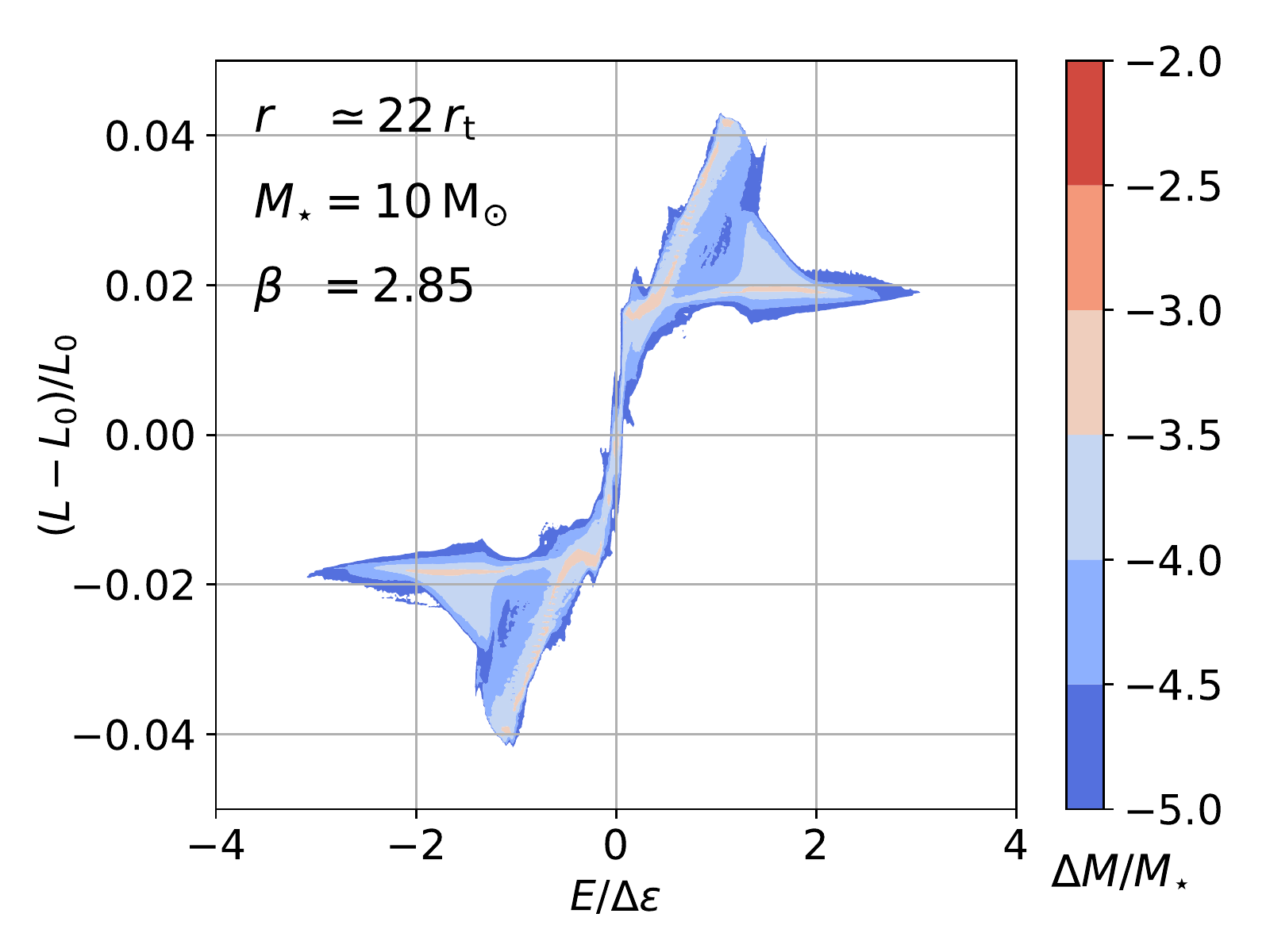}	
	\caption{The distribution of specific energy $E$ and specific angular momentum $L$ for $M_{\star}=0.3$ (\textit{top}), $1.0$ (\textit{middle}) and $10$ (\textit{bottom}).
	We consider the strongest encounter (largest $\beta$) for each star. The color scale indicates the mass fraction $\Delta M /M_{\star}$ in a logarithmic scale. We normalize $E$ by the fiducial energy spread $\Delta \epsilon$ (Equation~\ref{eq:deltae}) and $L_{0}$ refers to the initial specific angular momentum. 
	}
	\label{fig:dmdl}
\end{figure}

\subsection{Distribution of specific energy and angular momentum and fallback rate}
\label{subsub:dmde_ful}

The distribution of mass with energy and angular momentum determines both the orbits of tidal debris and the rate at which mass returns to the vicinity of the black hole.
Their joint distribution $d^2M/dEdL$ is presented in Figure~\ref{fig:dmdl}
for the 
debris of stars with $M_{\star}=0.3$ (\textit{top}), $1$ (\textit{middle}) and $10$ (\textit{bottom}); in each case, we show data from the smallest $r_{\rm p}$ we simulated. 
Here, $E$ is the relativistic specific energy in the black hole frame minus the rest mass energy, corresponding to the classical orbital energy and $L$ is  the relativistic specific angular momentum of the debris when they are expelled from the computational domain. We normalize $E$ to $\Delta \epsilon$, which is defined by \citep{Lacy1982,Rees1988},
\begin{align}
\Delta \epsilon &= \frac{GM_{\rm BH}R_{\star}}{r_{\rm t}^{2}},\\
& = 2.1 \times 10^{-4}M_{\star}^{2/3}R_{\star}^{-1}\left(\frac{M_{\rm BH}}{10^{6}}\right)^{1/3}c^2,\\
&= 1.9 \times 10^{17}M_{\star}^{2/3}R_{\star}^{-1}\left(\frac{M_{\rm BH}}{10^{6}}\right)^{1/3}\erg \gram^{-1}. 
\label{eq:deltae}
\end{align}
The $y-$axis in Figure \ref{fig:dmdl} indicates the difference between $L$ 
and the initial angular momentum $L_{0}$. 
Measured in units of $r_{\rm g}c$ for $M_{BH}= 10^6\Msol$, $L_{0}\simeq6.85$ for $M_{\star}=0.3$, $\simeq6.51$ for $M_{\star}=1$, and $\simeq9.49$ for $M_{\star}=10$.

The distributions in Figure~\ref{fig:dmdl} are, in all cases, very nearly symmetric around the origin with respect to both $E$ and $L$.  However, the ranges of both $E$ and $L$, when measured in terms of $\Delta \epsilon$ and $L_0$, are functions of stellar mass.   To characterize the width of these distributions, we define $\Delta E $ and $\Delta L$ such that 90\% of the total mass is contained within $-\Delta E < E < +\Delta E$ and $-\Delta L < L-L_0 < +\Delta L$.  The range of pink--red color in the figure is a good estimator of both $\Delta E/\Delta\epsilon$ and $\Delta L/L_{0}$.

\begin{figure}
	\centering
	\includegraphics[width=9.1cm]{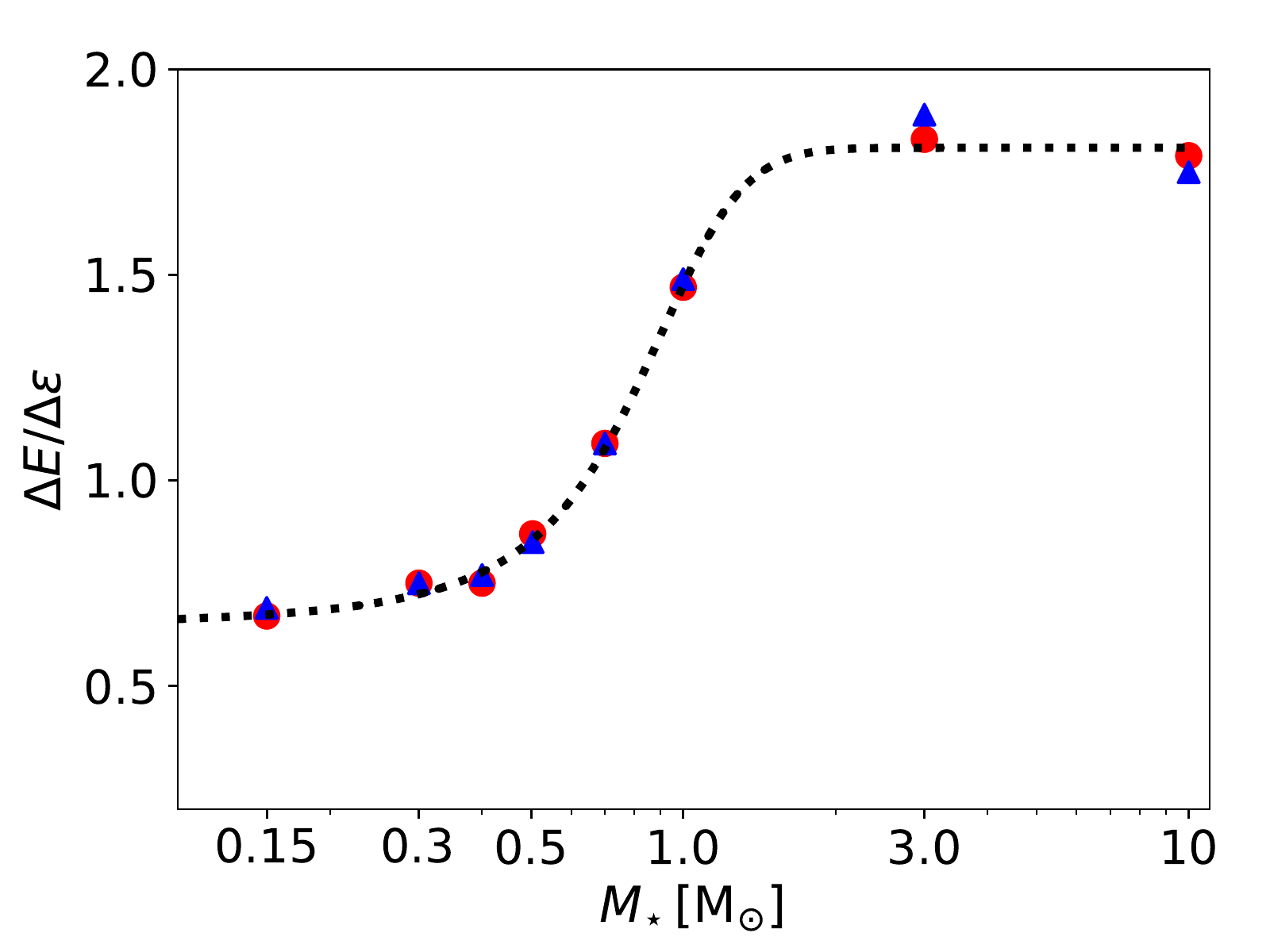}
	\caption{$\Delta E/\Delta\epsilon$ for all full disruption events ($r_{\rm p}<\mathcal{R}_{\rm t}$). When we have data for two values of $r_{\rm p} < \mathcal{R}_{\rm t}$ (see Table~\ref{tab:modelparameter}), the red circles indicate the smaller $r_{\rm p}$, while the blue triangles indicate the larger. The black dotted line represents the fitting formula for $\Delta E/\Delta\epsilon$ (Equation~\ref{eq:max_fit_DEdeps}). }
	\label{fig:E_width}
\end{figure}

Much as we found for $\mathcal{R}_{\rm t}$, there are strong contrasts between low-mass and high-mass stars for both $\Delta E/\Delta\epsilon$ and $\Delta L/L_{0}$. As $M_{\star}$ increases, $\Delta E/\Delta\epsilon$ jumps from $0.6 - 0.8$ to $1.8$ between $M_{\star}=0.5$ and $M_{\star}=3$ (see Figure~\ref{fig:E_width}).  In contrast, $\Delta L/L_{0} \approx 0.01$ 
for all $M_{\star} \leq 1$, but leaps to $\approx 0.02 - 0.04$ for higher masses.
As demonstrated in Figure~\ref{fig:E_width}, the value of $\Delta E/\Delta \epsilon$ is essentially unchanged over the $\approx 10 - 20\%$ span of pericenters inside $\physrad$ probed by our simulations.  Because $\Delta\epsilon$ is a function of $\rtidal$, but not $r_{\rm p}$, $\Delta E$ is also unchanged for pericenters close inside $\physrad$.  Such a weak dependence on $\beta$ is consistent with the Newtonian simulations of \citet{Guillochon+2013}; it is possible that for larger $\beta$ or larger $M_{\rm BH}$ relativistic effects could cause the energy spread to vary with $r_{\rm p}$.

The $M_{\star}$-dependence of $\Delta E/\Delta\epsilon$ is well-described by a fitting formula introduced in \citetalias{Ryu1+2019} (where it is called $\Xi_{\star}$),
\begin{align}
\frac{\Delta E(M_{\star})}
{\Delta \epsilon} =& \frac{0.620+\exp{[(M_{\star}-0.674)/0.212]}}{1 + 0.553~\exp{[(M_{\star}-0.674)/0.212]}}.
\label{eq:max_fit_DEdeps}
\end{align}

\begin{figure}
	\centering
	\includegraphics[width=9.1cm]{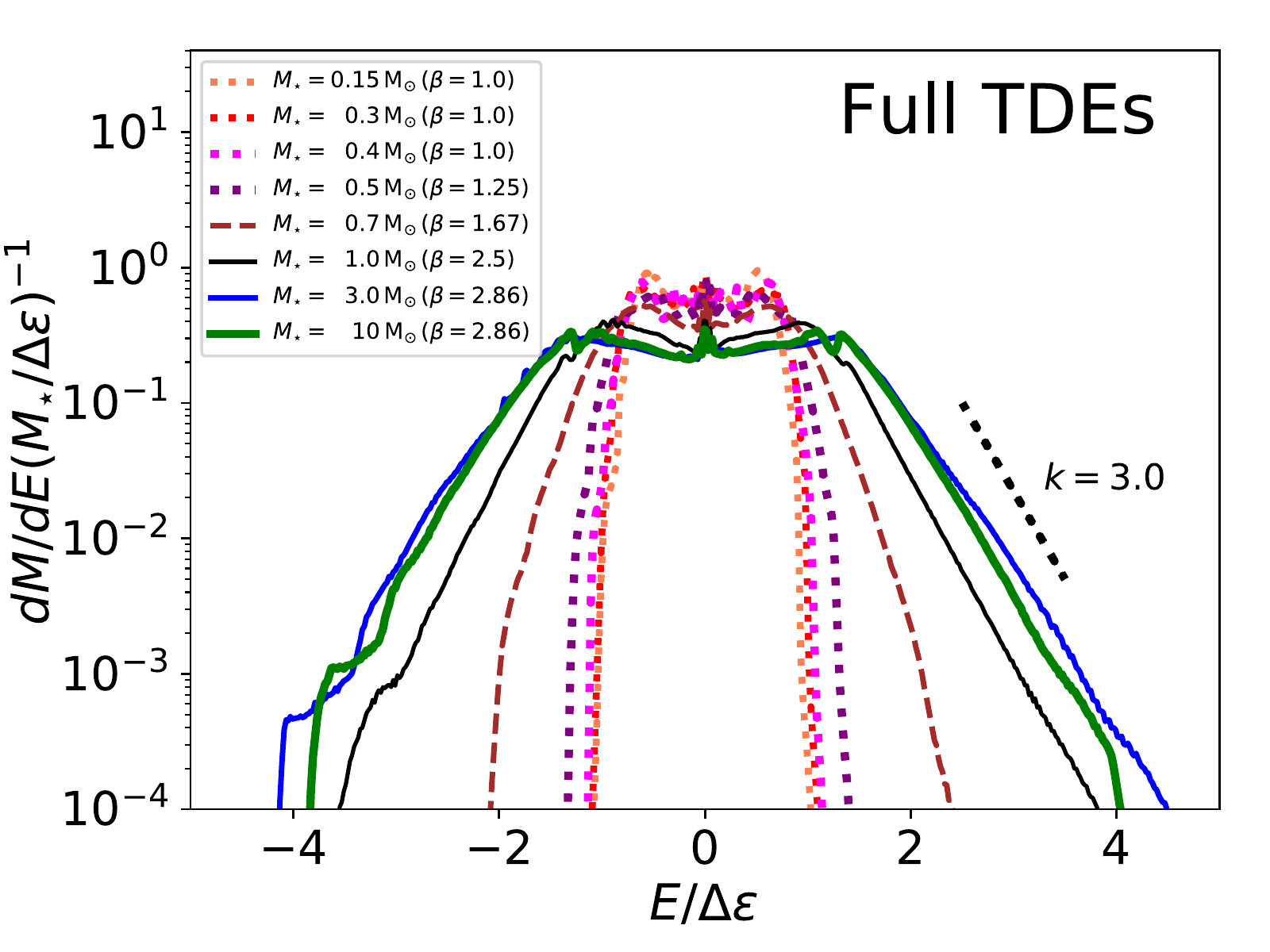}
	\caption{$dM/dE$ for the strongest encounter for each value of $M_{\star}$. These distributions are normalized so that the integrated area under the curve is unity. The diagonal dotted line indicates the slope if $dM/dE\propto e^{-k |E|/\Delta\epsilon}$ with $k=3.0$.}
	\label{fig:dmde}
\end{figure}

Figure \ref{fig:dmde} depicts $dM/dE$ for all of our fully-disrupted stars. As already mentioned,  the energy spread for high-mass stars is close to a factor of 2 broader than for low-mass ones when measured in terms of $\Delta\epsilon$.  Because $\Delta\epsilon \propto R_{\star}/r_{\rm t}^2$, this unit of energy is $\propto M_{\star}^{-0.45}$.

Although $dM/dE$ does not vary by large factors within its central region, neither is it strictly flat, as is often assumed.   For both low-mass and high-mass stars, the distribution has ``shoulders'', larger $dM/dE$ for $|E|/\Delta\epsilon \lesssim 1$ than for $E/\Delta\epsilon \simeq 0$.  The value of $dM/dE$ at the peaks of the shoulders is typically $\approx 1.5\times$ $dM/dE$ at the local minimum near $E=0$.  The distribution has fairly sharp outer boundaries for the low-mass stars, but a more gradual fall for the high-mass stars. Where $|E| > \Delta E$, $dM/dE$ is very well described by an exponential $\exp[{-k |E|/\Delta\epsilon}]$. For $M_{\star} < 0.7$, $k \gtrsim 7$, but $k$ falls to $\simeq 2.5$--3.0 for $M_{\star} \geq 1$.

The spikes at $E \simeq0$ represent the last remaining gas in the simulation box. As the remnant moves farther out, both the width of this spike and the integral under it decrease. These features are also reported in other studies \citep[e.g][]{Lodato+2009,Coughlin+2016}.

\begin{figure}
	\centering
	\includegraphics[width=9.1cm]{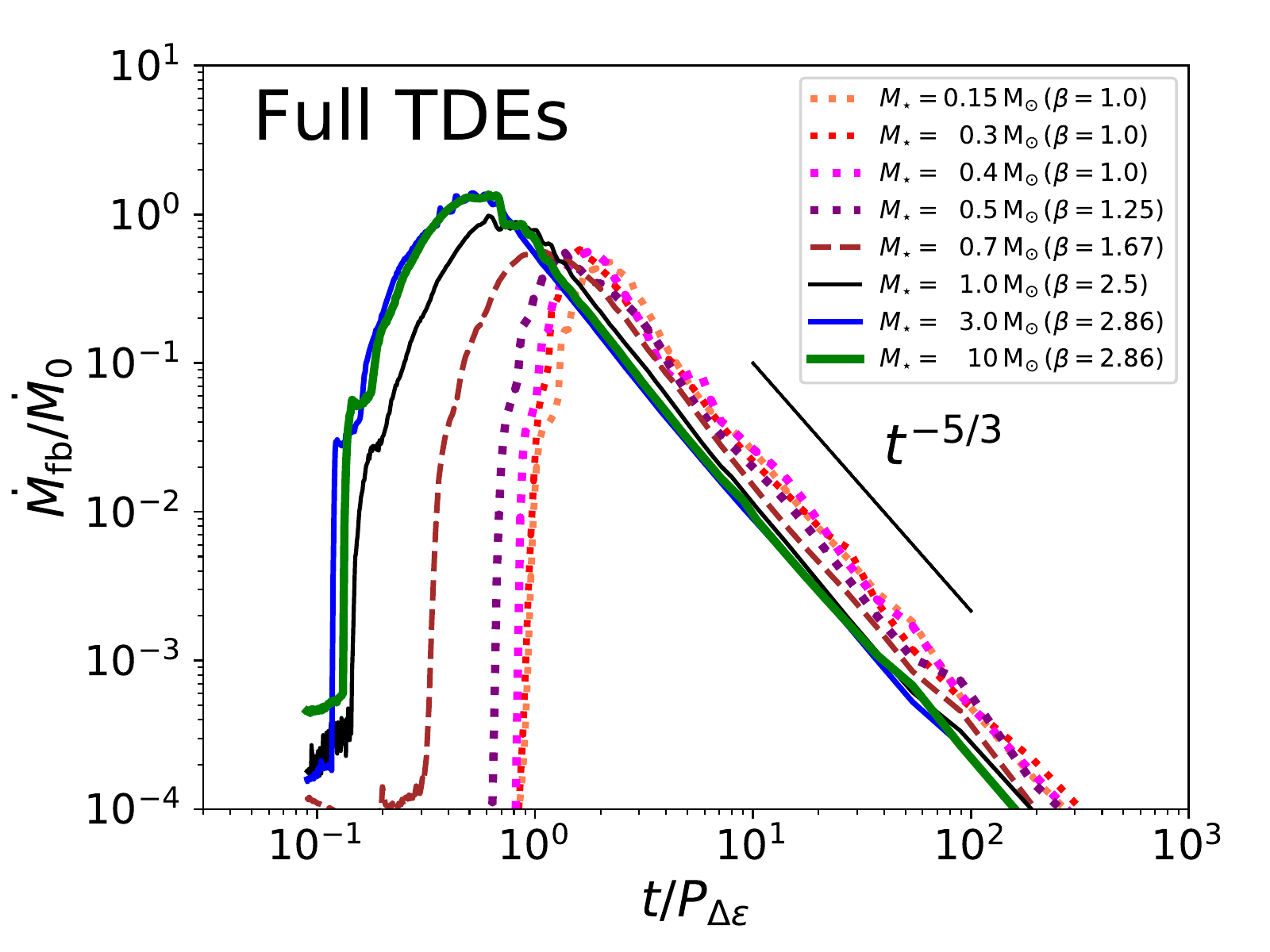}	
	\caption{The fallback rate $\dot{M}_{\rm fb}$ for the same TDEs shown in Figure~\ref{fig:dmde}. We normalize the time $t$ by the orbital period $P_{\Delta\epsilon}$
	and the fallback rate by $\dot{M}_{0}$. The diagonal solid line shows the conventional power-law $t^{-5/3}$. }
	\label{fig:fallback}
\end{figure}

Using the energy distribution data from our simulations (Figure~\ref{fig:dmde}) and the expression for the fallback rate \citep{Rees1988,Phinney1989},
\begin{align}
\dot{M}_{\rm fb}=\left(\frac{M_{\star}}{3P_{\Delta\epsilon}}\right)\left( \frac{dM/M_{\star}}{dE/2\Delta\epsilon}\right)\left(\frac{t}{P_{\Delta\epsilon}}\right)^{-5/3},
\label{eq:mdot2}
\end{align} 
 we determine the fallback rate (see Figure~\ref{fig:fallback}).
It is useful to define two normalization scales: $P_{\Delta\epsilon}= (\uppi/\sqrt{2})G M_{\rm BH}\Delta\epsilon^{-3/2}$, the orbital period for orbital energy $-\Delta\epsilon$; and $\dot{M}_{0}=M_{\star}/(3P_{\Delta\epsilon})$, the characteristic mass-return rate.

For  full disruptions, the shapes of the fallback rate curves divide neatly into two classes, as expected from the distinctive shapes of the energy distributions. For low-mass stars, a steep rise that reaches a maximum fallback rate $\dot{M}_{\rm max}\simeq0.5\dot{M}_{0}$ at $t\simeq(1.5-2)P_{\Delta\epsilon}$ is followed by a quick transition to a $t^{-5/3}$ decay.  On the other hand, because the energy spread $\Delta E$ for the most-bound debris from high-mass stars is $\approx 2\Delta\epsilon$, the fallback rate for these stars peaks earlier, at $t\simeq0.5 P_{\Delta\epsilon}$, and at a higher rate, $\dot{M}_{\rm max}\simeq(0.8-1.3)\dot{M}_{0}$. The return rate of the stellar debris from $0.7\Msol$ stars lies between that of low-mass and high-mass stars.

\section{A single semi-analytic model for both physical tidal radius and remnant mass}
\label{subsec:analyticmodel}

We have shown that the traditional order-of-magnitude model for tidal radii needs to be corrected with order-unity coefficients in order to match quantitatively the behavior of realistic main sequence stars. Here we show how a natural generalization of the original tidal radius argument, augmented by a single free parameter, can be used both to deepen our understanding of the order-unity coefficients and to predict how much mass is lost in a partial disruption. A qualitative version of this argument was made by \citet{Li+2002}, but was never applied to actual stellar structures.

Suppose that the amount of mass stripped from a star during the entire event is the mass in the unperturbed star outside the radius such that the star's self-gravity at that location is a factor $\zeta$ times the tidal force applied at that radius when the star is at pericenter.  In other words,
\begin{align}
\frac{GM(R)}{R^{2}} = \zeta \frac{GM_{\rm BH}R}{r_{\rm p}^{3}},
\label{eq:semi_condition}
\end{align}
where $M(R)$ is the enclosed mass inside $R$.
Replacing $r_{\rm p}$ with $\beta^{-1}~ r_{\rm t}$ and using the definition of $\rtidal$, Equation~\ref{eq:semi_condition} becomes
\begin{align}
\zeta^{-1}\beta^{-3}\left[\frac{M(R)}{M_{\star}}\right]\left[\frac{R_{\star}^{3}}{R^{3}}\right]=1.
\label{eq:analytic_condition}
\end{align}
Defining $\rho_{\star}=3M_{\star/}(4\pi R_{\star}^{3})$ and $\bar{\rho}(R)= 3M(R)/(4\pi R^{3})$ we finally have
\begin{align}
\beta^{-1}=\frac{r_{\rm p}}{r_{\rm t}}=\left[\zeta\left(\frac{\bar{\rho}_{\star}}{\bar{\rho}(R)}\right)\right]^{1/3}.
\label{eq:analytic:psi}
\end{align}
Thus, for a given pericenter distance $r_{\rm p}$ and density profile, we 
have an implicit solution for the radius $R$ beyond which the mass of the star is lost due to tidal forces. The enclosed mass $M(R)$ at the radius $R$ corresponds to the remnant mass.

\begin{figure}
	\centering
	\includegraphics[width=9.1cm]{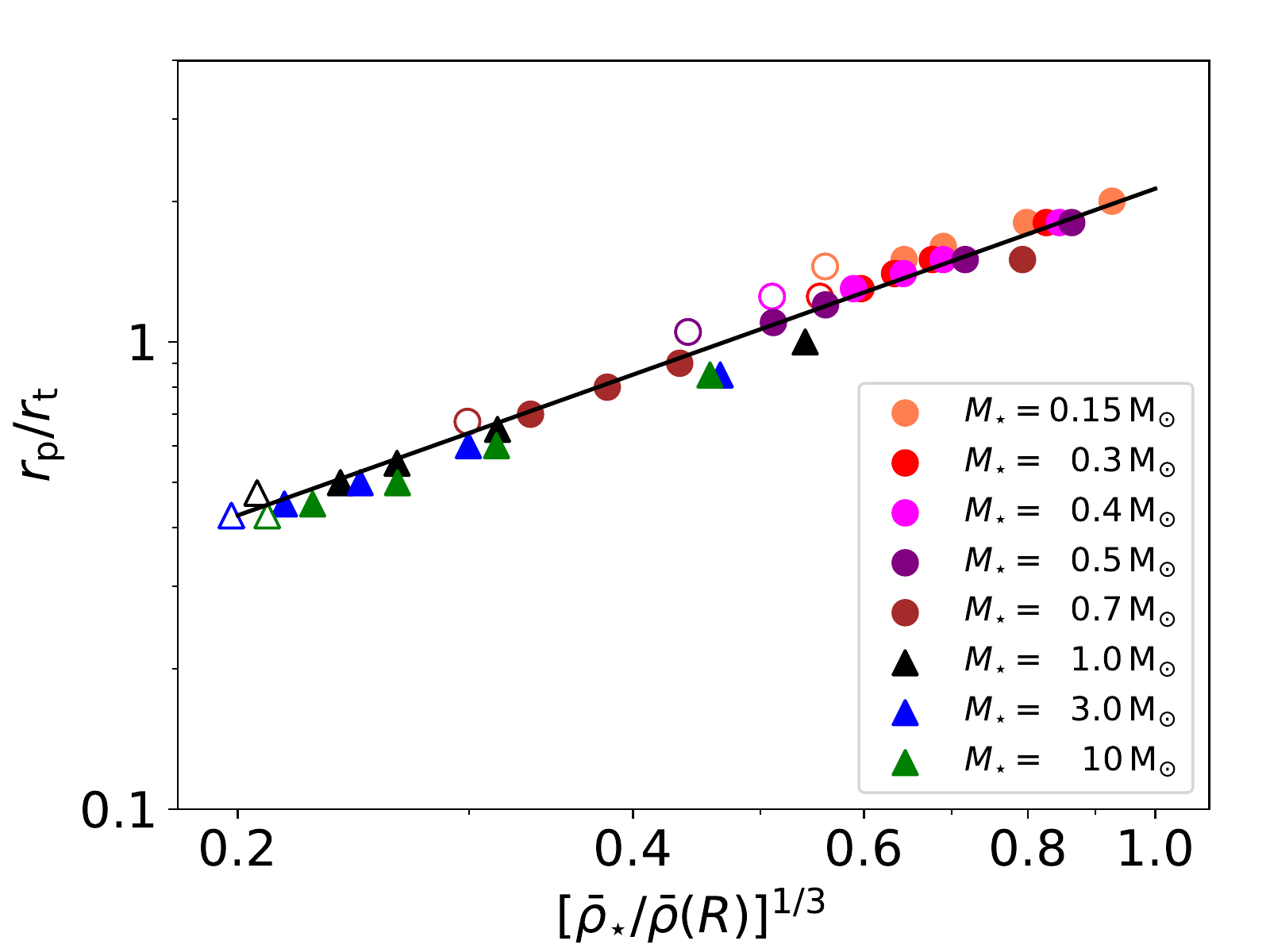}	
	\caption{Correlation between the density ratio $[\bar{\rho}_{\star}/\bar{\rho}(R)]^{1/3}$ and the corresponding pericenter.  Partial disruptions are shown with filled symbols, full disruptions with hollow.  The solid diagonal line is our best-fit linear model.}
	\label{fig:analyticmodel}
\end{figure}

In  searching for $\mathcal{R}_{\rm t}$, we ran simulations for numerous partial disruptions with varying $r_{\rm p}$ and studied the properties of the partially disrupted stars including the remnant mass. We will discuss our results in detail in \citetalias{Ryu3+2019}, but here we merely use the results. Using the remnant mass from the partial disruption simulations, we use the {\small MESA}~ enclosed mass profile $M(R)$ for each star to find $R$ such that the enclosed mass equals the remnant mass for that case.  We then compute the density ratio of Equation~\ref{eq:analytic:psi}.  The data shown in Figure~\ref{fig:analyticmodel} are the result.  The black line shows the best fit {\it assuming} that the relationship is linear; the figure makes it plain that this assumption is well-supported by the data.  The coefficient $\zeta \simeq 9.8$.  Thus, the remnant mass produced when a star passes a black hole with a given pericenter outside $\mathcal{R}_{\rm t}$ can be easily determined by use of \mesa~ models for the original structure of the star.

The limit of $R \rightarrow 0$ corresponds to a complete disruption. In that case, $\bar{\rho}=\rho_{\rm c}=\lim\limits_{R\rightarrow0}\rho(R)$. In other words, $\Psi$ can be determined solely from the ratio between the star's central density $\rho_{\rm c}$ and its mean density $\bar{\rho}_{\star}$:
\begin{align}
\Psi\simeq \left[\zeta\left(\frac{\bar{\rho}_{\star}}{\rho_{\rm c}}\right)\right]^{1/3}. 
\label{eq:Psi_analytic}
\end{align}
It follows that, unlike the traditional tidal radius $\rtidal$, which depends on the star's {\it mean} density,  $\mathcal{R}_{\rm t}$ is determined solely from its {\it central} density $\rho_{\rm c}$:
\begin{align}
\mathcal{R}_{\rm t} &= \Psi \rtidal \simeq {\zeta}^{1/3}\left(\frac{\bar{\rho}_{\star}}{\rho_{\rm c}}\right)^{1/3}\rtidal,\nonumber\\
&\simeq\left( \frac{3\zeta}{4\pi}\right)^{1/3} \left(\frac{M_{\rm BH}}{\rho_{\rm c}}\right)^{1/3},
\label{eq:Rt-zeta}
\end{align}
where $ [3\zeta/(4\pi)]^{1/3} \simeq 1.32$. This is the argument underlying Equation 6 in \citetalias{Ryu1+2019}.

The opposite limit, the pericenter distance outside which no mass is lost, is also instructive.  It is obtained by considering our underlying assumption, expressed by Equation~\ref{eq:semi_condition} and vindicated by Figure~\ref{fig:analyticmodel}, in the limit $R\rightarrow R_{\star}$, at which $r_{\rm p} \rightarrow \widehat R_{\rm t}$, the largest pericenter for any sort of tidal mass-loss.  In this limit, the equation takes the form
\begin{align}
    \widehat{R}_{\rm t}& = \zeta^{1/3} \rtidal \simeq 2.1 \rtidal.
\end{align}
Thus, the ratio $\widehat{R}_{\rm t}/r_{\rm t}$ has no explicit dependence on $M_{\star}$.  In addition, it is the limiting pericenter for partial disruptions that depends most closely on the star's mean density.  It should, perhaps, not be surprising that it is only the ability to remove a small amount of matter from the outside of the star that depends on the competition between tidal gravity and self-gravity at the star's edge.

This limit may also be described in a different way.  Dividing Equation~\ref{eq:analytic:psi} by $\Psi$ using Equation~\ref{eq:Psi_analytic} yields
\begin{align}
\label{eq:analytic_interme}
\frac{r_{\rm p}}{\mathcal{R}_{\rm t}}=\left[\frac{\rho_{\rm c}}{\bar{\rho}(R)}\right]^{1/3}.
\end{align}
The maximum pericenter for losing any mass is then
\begin{align}
\widehat R_{\rm t} = \left(\frac{\rho_c}{\bar\rho_{\star}}\right)^{1/3}\mathcal{R}_{\rm t}.
\end{align}
In other words, the ratio between the maximum pericenter for a partial disruption and the maximum pericenter for a full disruption increases with the degree of central concentration $\rho_{\rm c}/\bar{\rho}_{\star}$.  It is therefore larger for high-mass stars than for low-mass.

\section{Discussion: comparison with previous studies}
\label{sec:discussion}

\subsection{Physical tidal radius}
\label{subsub:comparison_rt}

\begin{table*}
	\centering
	\caption{Examples of a Previous Study in Which Characteristic Tidal Distances Are Identified. Notes: The first row shows our result for $M_{\star}=1$. We list their numerical methods (second column), stellar models (third column), $\mathcal{R}_{\rm t}/r_{\rm t}(\equiv\Psi)$ (fourth column).}
	\begin{tabular}{c   c c  c c }
		\hline\hline\noalign{\smallskip}
		Reference  &   method/code  & stellar model   & \multicolumn{2}{c}{$\mathcal{R}_{\rm t}/r_{\rm t}(\equiv\Psi)$}  \\
		\hline\noalign{\smallskip}
		This work    & GRHD/HARM3D$^{a}$ & $1\Msol$ (\mesa) & \multicolumn{2}{c}{$0.475\pm0.025$} \\
			\hline\noalign{\smallskip}
		\citealt{Phinney1989}    & $\cdot\cdot\cdot$ & $\cdot\cdot\cdot$ & $0.82^{b}$    & $0.52^{c}$  \\
	\noalign{\smallskip}	\hline\noalign{\smallskip}
		\citealt{Guillochon+2013}  &  AMR$^{f}$ & $1\Msol$, $1\Rsol$ (Polytrope)& $1.11^{d}$ &$0.54^{e}$  \\
	\noalign{\smallskip}	\hline\noalign{\smallskip}
		\citealt{Mainetti+2017} & AMR, MFM$^{g}$, SPH$^{h}$   & $1\Msol$, $1\Rsol$ (Polytrope)&  $1.08^{d}$  &$0.50^{e}$    \\
	\noalign{\smallskip}	\hline\noalign{\smallskip}
		\citealt{Goicovic+2019}& MM$^{i}$   & $1\Msol$ (\mesa, ZAMS$^{j}$)& \multicolumn{2}{c}{$0.50$} \\
	\noalign{\smallskip}	\hline\noalign{\smallskip}
		\multirow{4}{*}{\citealt{Law-Smith+2020}}& \multirow{4}{*}{AMR}   & $0.15\Msol$ (\mesa, MAMS$^{k}$)& \multicolumn{2}{c}{$0.11$} \\
              &  & $0.5\Msol$ (\mesa, MAMS$^{k}$)& \multicolumn{2}{c}{$0.91$} \\
	&    & $0.7\Msol$ (\mesa, MAMS$^{k}$)& \multicolumn{2}{c}{$0.63$} \\
	&    & $1.0\Msol$ (\mesa, MAMS$^{k}$)& \multicolumn{2}{c}{$0.37$} \\
	\noalign{\smallskip}	\hline\noalign{\smallskip}
		\multicolumn{5}{l}{$^{a}$ General relativistic magneto-hydrodynamics; $^{b}$ fully-convective stars; $^{c}$ fully-radiative stars; }\\
		\multicolumn{5}{l}{ $^{d}$ polytropic model with $\gamma=5/3$; $^{e}$ polytropic model with $\gamma=4/3$; $^{f}$ AMR: Adaptive mesh refinement;  }\\
		\multicolumn{5}{l}{$^{g}$ MFM: mesh-free finite mass; $^{h}$ SPH: smoothed particle hydrodynamics; $^{i}$ MM: moving mesh}\\
		\multicolumn{5}{l}{$^{j}$ ZAMS : zero-age main sequence star; $^{k}$ MAMS : middle-age main sequence star}
		\label{tab:previous}
	\end{tabular}
\end{table*}

 Figure~\ref{fig:tidal_distance}  compares our results for $\mathcal{R}_{\rm t}/r_{\rm t}(\equiv \Psi)$ with other simulations and with the correction factor introduced by \citet{Phinney1989}. We also tabulate the results from other numerical studies in Table~\ref{tab:previous}.

The dramatic change in $\Psi$ from $M_{\star}=0.4$ to $M_{\star}=1$ is due to change in the internal structure of the stars. This trend was predicted by \citet{Phinney1989}, who suggested adjusting $r_{\rm t}$ by the factor $(k/f)^{1/6}$, in which $k$ is the apsidal motion constant, reflecting the degree of central concentration, and $f$ is the non-dimensional binding energy. Low-mass stars, which are convective except possibly near their core, tend to be rather less centrally concentrated than high-mass stars, which are convective only near their cores (see Figure~\ref{fig:densityprofile}).  Phinney's model leads to a prediction that $\Psi_{k/f} = 0.82$ for fully-convective stars  (e.g., $0.15-0.4\Msol$) and $\Psi_{k/f} = 0.52$ for fully-radiative stars (e.g., $1\Msol$). The qualitative sense of this prediction is consistent with our results ($\Psi = 1.25-1.45$ for $M_{\star} \leq 0.3$ and $\Psi = 0.425$ for $M_{\star} \geq 3$). 

Earlier numerical simulations of TDEs 
\citep{Guillochon+2013,Mainetti+2017} approximated MS stars by polytropic models.
\citet{Guillochon+2013} focused on the mass fallback rate, using the adaptive-mesh refinement (AMR) grid-based hydrodynamics code {\small FLASH}. They considered  only $M_{\star}=1$ with $\gamma=4/3$ and $5/3$ and assumed that a star is completely disrupted when the logarithmic time derivative of the self-bound stellar mass remains $\sim O(1)$ for all times after the time of pericenter passage. With this definition, they found that $\Psi \simeq 0.54$ for $\gamma=4/3$ and $\simeq1.1$ for $\gamma=5/3$.
\citet{Mainetti+2017} measured $\Psi$ using three numerical techniques: mesh-free finite mass, smoothed particle, and AMR grid-based hydrodynamics simulations; they then checked that the different techniques gave consistent results.  Likewise considering polytropic stars with the same values of $\gamma$ and a similar disruption criterion, they found results very close to those of \citet{Guillochon+2013}: $\Psi\simeq0.5$  for $\gamma=4/3$ and $\simeq1.08$  for $\gamma=5/3$. 
For our fully-convective stars, those with $M_{\star}=0.15-0.4$, we find a physical tidal radius larger by 15--30\%, $\Psi\simeq1.25-1.45$. It is very likely that this contrast is due to our use of fully relativistic tidal stresses because we find $\Psi = 1.15\pm0.05$ for a fully-convective star when $M_{\rm BH} = 10^5$, and $\Psi$ increases for larger $M_{\rm BH}$ as would be expected for a relativistic effect (\citetalias{Ryu4+2019}). 
For $M_{\star}=1$, a polytrope with index corresponding to $\gamma = 4/3$ coincidentally gives a fairly good approximation to the actual density profile (see Figure \ref{fig:densityprofile}); at this mass, we find $\Psi=0.475$ , 14\% less than the value found from the Newtonian polytropic assumption (and the $(k/f)^{1/6}$ prediction).
However, this offset must be due to the actual structural contrast, not relativistic effects, because it is even larger for smaller black hole mass: $\Psi \simeq 0.425\pm0.05$ for $M_{\rm BH} = 10^5$ (\citetalias{Ryu4+2019}).  At higher masses, the $\gamma = 4/3$ Newtonian polytrope approximation becomes still poorer, overestimating $\Psi$ by 27 \% for $M_{\star} = 1$.

Most recently, several studies using \mesa~  to create the initial stellar model have been published. \citet{Goicovic+2019} performed hydrodynamics simulations for TDEs of a $M_{\star} =1$ zero-age main sequence star using the moving-mesh code {\small AREPO}. Their definition of full disruption was that of \citet{Guillochon+2013}. They found $\Psi=0.5$, essentially in agreement with the polytropic-model calculations of \citet{Guillochon+2013} and \citet{Mainetti+2017}. \citet{Law-Smith+2020} performed hydrodynamics simulations using the AMR code {\small FLASH} based on \mesa~models for main-sequence stars at zero-age, middle-age, and terminal-age; for the middle-age $M_{\star}=1$ case, they found $\Psi\simeq0.37$.

Thus, where our results pertain to the same stellar model, they agree qualitatively with previous work, but with two interesting discrepancies. For the case of $M_{\rm BH} = 10^6$ presented here, full tidal disruptions can occur for larger pericenters than previously thought.  As we will analyze more carefully in \citetalias{Ryu4+2019}, this discrepancy can  be attributed to relativistic effects that only we have included.   Because the relativistic effects strengthen with increasing $M_{\rm BH}$, the differences can be substantial when $M_{\rm BH} > 10^6$. Secondly, for middle-aged main-sequence stars with $M_{\star} \gtrsim 0.5$, the polytropic approximation is quantitatively inadequate.



\subsection{Debris energy distribution}\label{subsub:energydebris}

Only two previous papers presented details of the $dM/dE$ distribution.
\citet{Guillochon+2013} studied $\gamma=5/3$ and $\gamma=4/3$ polytropic models for $M_{\star}=1$; for the latter case, the one more appropriate to stars of this mass, the energy associated with the peak of mass-return was, in our notation, $\simeq 1.3\Delta\epsilon$.  Because \citet{Goicovic+2019}, who used a MESA internal density profile, presented plots, but no numerical values, their result appears to be equally consistent with both that of \citet{Guillochon+2013} and our value, $\simeq 1.5\Delta\epsilon$.

The energy distribution figure displayed by \citet{Goicovic+2019} also shows exponential wings like our $dM/dE$, and with an approximately similar slope.

\section{Summary}
\label{sec:summary}

This is the second installment in a series of papers reporting on our program of tidal disruption simulations in which the stars are given realistic main-sequence internal structures, and the gravitational dynamics are treated in full general relativity.

In our first paper (\citetalias{Ryu1+2019}), we presented an overview and highlighted our results with the greatest observational implications.  Here we described the details of our calculations and our findings regarding events in which the stars are completely disrupted by a $10^{6}\Msol$ BH.

Our calculations are noteworthy in several respects: their fully relativistic treatment of dynamics due to the black hole's gravity; their employment of \mesa~ to determine the initial conditions, so that they begin with density profiles of realistic stars; and the large range of stellar masses explored and the relatively dense coverage of that mass-range, properties that enable us to clearly determine how mass-dependence modifies the order of magnitude picture.   Although in this work we present results for a SMBH of $10^6\Msol$, in \citetalias{Ryu4+2019} we also explore the black hole mass-dependence of these correction factors.

Previous work employing Newtonian dynamics had noted that the physical tidal radius for polytropes with $\gamma=5/3$, a good model for fully-convective stars, is actually slightly greater than the widely-used order-of-magnitude estimate $\rtidal \equiv R_{\star} (M_{\rm BH}/M_{\star})^{1/3}$, while the physical tidal radius for a polytrope with $\gamma=4/3$, a coincidentally good match to stars of mass $M_{\star}=1$, but not to any others, is $\gtrsim 0.5\rtidal$.  We have shown that for fully-convective stars ($M_{\star} \leq 0.3$) encountering a black hole whose mass is $10^{6}$, the actual physical tidal radius is several tens of percent {\it greater} than the Newtonian prediction ($\simeq 1.4\rtidal$ rather than $\simeq 1.1\rtidal$).  As demonstrated in \citetalias{Ryu1+2019} and \citetalias{Ryu4+2019}, this contrast is a relativistic effect.  We have further shown that for $M_{\star} \geq 3$, $\mathcal{R}_{\rm t} \simeq 0.4 \rtidal$.  There is a sharp (but continuous) transition between these two limits across the range of masses $M_{\star} = 0.5$--1.  For $M_{\rm BH} = 10^6$, the physical tidal radius of {\it all} stars with $0.15\leq M_{\star} \leq 3$ is $\simeq 27~r_{\rm g}$ to within $\pm 20\%$ (\citetalias{Ryu1+2019}).

In addition, we have demonstrated that although the characteristic debris energy scale suggested by \citet{Lacy1982} is a reasonable estimator of the actual width of the debris energy distribution, it requires factor $\sim 2$ corrections dependent upon the stellar mass.  Like the ratio between physical tidal radius and nominal tidal radius, these corrections are roughly constant as a function of stellar mass at both the high and low ends of the range, but these constants are different.  In addition, although the distribution of mass with energy has been widely assumed to be flat between sharp edges ever since the work of \citet{Rees1988} and \cite{EvansKochanek1989}, we have found that for all stars the distribution has ``shoulders" near $E \approx \Delta E$ at which $dM/dE$ is $\approx 50\%$ greater than $dM/dE$ at $E=0$, where there is a local minimum.  Moreover, although the edges of the distribution for fully-convective stars are, indeed, quite sharp, the energy distribution for debris from stars with $M_{\star} \geq 1$ generically has wings containing a small, but possibly significant amount of mass with energy 2--$3\Delta E$.

These results strengthen the critical questions raised by the popular ``frozen-in" approximation.  In its most ambitious form \citep{Lodato+2009,Stone+2013}, it has been used to predict the ultimate energy distribution of the debris based entirely on the matter's potential energy within the undisturbed star at radii close to the black hole (sometimes $\rtidal$, sometimes $\physrad$, sometimes $r_{\rm p}$).  In particular, we have shown that mass-loss begins only shortly after pericenter passage, and continues (in complete disruptions) until the star has reached a distance from the black hole $\approx 20\rtidal$, which can be  $\approx50~\mathcal{R}_{\rm t}$.  Throughout this entire time, the {\it instantaneous} tidal radius $\lambda_{\rm t} \sim r$. Thus, the specifics of the energy distribution are determined by continued interaction between the black hole's gravity, the star's self-gravity, and internal fluid forces.

Our estimates of the physical tidal radius affect, among other things, the rate of full TDEs, as well as the relative rates for stars of different masses. Our alterations to the expected energy distribution lead immediately to implications regarding the rate and time-delay at which matter falls back to the star. These changes are especially noteworthy for the more massive stars, as they predict a time of peak fallback several times earlier than the traditional prediction, and a maximum rate correspondingly larger.  As emphasized in \citetalias{Ryu1+2019}, these corrections can be important in any attempt to relate observed light curves to the fallback rate, and from the constraints obtained determine the system's parameters.

\section*{Acknowledgements}
We thank the anonymous referee for comments and suggestions that helped us to improve the paper.
This work was partially supported by NSF grant AST-1715032, Simons Foundation grant 559794 {and an advanced ERC grant TReX}. S.~C.~N. was supported by the grants NSF AST 1515982, NSF OAC 1515969, and NASA 17-TCAN17-0018, and an appointment to the NASA Postdoctoral Program at the Goddard Space Flight Center administrated by USRA through a contract with NASA.  The authors acknowledge the analysis toolkit matplotlib \citep{Hunter:2007} for making the plots in the paper. This research project (or part of this research project) was conducted using computational resources (and/or scientific computing services) at the Maryland Advanced Research Computing Center (MARCC). The authors would like to thank Stony Brook Research Computing and Cyberinfrastructure, and the Institute for Advanced Computational
Science at Stony Brook University for access to the high-performance
SeaWulf computing system, which was made possible by a $\$1.4$M National Science Foundation grant (\#1531492).

\software{
matplotlib \citep{Hunter:2007}; \mesa \citep{Paxton+2011}; 
\harm  \citep{Noble+2009}.}


\appendix
\section{Validity of calculating stellar self-gravity via the Poisson Equation}\label{appendix}

Our description of stellar self-gravity rests on two assumptions: that there exists a coordinate frame whose origin coincides with the star's center-of-mass and in which the metric is close to Minkowski throughout our problem volume; and that it is legitimate to calculate the stellar self-gravity without reference to any time-dependence it may have.  The first of these statements may be rephrased as stating the  metric in this frame can be written as
\begin{align}\label{eq:metric}
g_{\mu\nu} \simeq \eta_{\mu\nu}+ h^{\rm tidal}_{\mu\nu} + h_{\mu\nu}^{\rm sg},
\end{align}
where $\eta_{\mu\nu}$ is the Minkowski metric and both $|h_{\mu\nu}^{\rm tidal}|$ and $|h_{\mu\nu}^{sg}|$ are $\ll 1$.  Here $\eta_{\mu\nu}+h^{\rm tidal}_{\mu\nu}$ is the global Schwarzschild metric after a coordinate transformation into this frame.  The metric perturbation $h_{\mu\nu}^{\rm sg}$ is due to the star's self-gravity; because its magnitude is small, we assume all its components are zero except $h_{00}^{\rm sg}=-2 \Phi_{\rm sg}/c^{2}$, where $\Phi_{\rm sg}$ is the star's potential. In this Appendix, we will explicitly estimate the parameter bounds for which these assumptions are justified and demonstrate how they validate our choice to compute $\Phi_{\rm sg}$ from the Poisson equation. Throughout this Appendix, Greek indices (e.g., $\mu$, $\nu$, $\lambda$) run over the four coordinate labels ($t$, $x$, $y$, $z$), while Latin indices (e.g., $i$, $j$) refer to spatial coordinate labels ($x$, $y$, $z$). We apply Einstein summation notation only to Greek indices.

The contribution of stellar self-gravity is $\sim 10^{-6}$, and is therefore always sufficiently small.
As explained in \ref{subsec:selfgravity}, to satisfy the condition $|h_{\mu\nu}^{\rm tidal}| \ll 1$ we construct an orthonormal tetrad basis for the comoving frame.  This procedure guarantees that $h^{\rm tidal}_{\mu\nu} \equiv 0$ at the origin. When the radial coordinate of the box origin is $\gg r_{\rm g}$, it also results in $|h^{\rm tidal}_{\mu\nu}| \ll 1$ throughout the box.  However,  $|h^{\rm tidal}_{\mu\nu}|$ grows as the separation between the star and the BH decreases. In most of the box's volume, $|h^{\rm tidal}_{\mu\nu}| \simeq 10^{-4}$ for $r_{\rm p}/r_{\rm g}\simeq 100$, $\simeq 10^{-3}$ for $r_{\rm p}/r_{\rm g}\simeq 20$, and $\simeq10^{-2}$ for $r_{\rm p}/r_{\rm g}\simeq 5$.  To describe it as a function of $r_{\rm p}/\rg$, we fit our data on $h^{\rm tidal}_{\mu\nu}$ with the form $\lambda \left(r_{\rm p}/r_{\rm g}\right)^{-n}$, finding $n\simeq1.6$ and $\lambda \simeq 0.2$ in the majority of the domain volume. Thus, the tidal terms are, indeed, small provided that $r_{\rm p}/\rg \gtrsim 10$; for the largest black hole mass we treat, $M_{\rm BH} = 10^7$, the physical tidal radius ${\cal R}_{\rm t} \simeq 9~\rg$.

 Whether these values of $|h_{\mu\nu}^{\rm tidal}|$ are sufficiently small that our Poisson equation calculation of $\Phi_{\rm sg}$ is accurate depends on a different criterion: whether these small perturbations might lead to terms in the Einstein Field Equations, the true gravitational field equations, large enough to alter $\Phi_{\rm sg}$ substantially. To test our method against this criterion, we will perform a perturbative expansion of the Einstein field equations in terms of $h_{\mu\nu}^{sg}$ and $h_{\mu\nu}^{\rm tidal}$.
For this purpose, it is convenient to write the field equations in the form \citep{Weinberg1972},
\begin{align}\label{eq:field}
    R_{\mu\nu}=-8\uppi G (T_{\mu\nu}-\frac{1}{2}g_{\mu\nu}T^{\lambda}_{~~\lambda}),
\end{align}
where $R_{\mu\nu}$ is the Ricci tensor and $T_{\mu\nu}=\rho h u^{\mu}u^{\nu}+p g^{\mu\nu}$ is the stress-energy tensor. Here, $\rho$ is the  proper rest-mass density, $h$ is the enthalpy, $u^{\mu}$ is the fluid 4-velocity and $p$ is the pressure. 

For our purposes, the t--t component of this tensor equation is the most important because it is the only one relevant to $h_{\mu\nu}^{\rm sg}$. Expanding the portion of the Ricci tensor linear in the connections to show its explicit dependence on the metric and its derivatives, we find
\begin{align}\label{eq:R1}
    R_{tt} &=g^{\lambda\nu}\frac{1}{2} (\partial_{t}^{2}g_{\lambda\nu}-\partial_{t}\partial_{\lambda}g_{t\nu}-\partial_{t}\partial_{\nu}g_{t\lambda}+\partial_{\nu}\partial_{\lambda}g_{tt})
    +g^{\lambda\nu}g_{\eta\sigma}(\Gamma^{\eta}_{\lambda\nu}\Gamma^{\sigma}_{tt}-\Gamma^{\eta}_{t\lambda}\Gamma^{\sigma}_{t\nu})\\
    &\simeq \frac{1}{2}\sum_{i=1}^{3}g^{ii}(\partial_{t}^{2}g_{ii}-2\partial_{t}\partial_{i}g_{ti}) +\frac{1}{2}\nabla^{2}g_{tt}
    +g^{\lambda\nu} g_{\eta\sigma}(\Gamma^{\eta}_{\lambda\nu}\Gamma^{\sigma}_{tt}-\Gamma^{\eta}_{t\lambda}\Gamma^{\sigma}_{t\nu}),
\end{align}
where $\partial_{\nu}$ refers to the ordinary partial derivative with respect to $\nu$ and $\nabla^{2}$ is the spatial Laplace operator.  The second form results from the fact that all terms in the first bracket with $\lambda=t$ and $\nu=i$ or $\lambda=t$ and $\nu=t$ cancel each other, and the terms with $\lambda=i$ and $\nu=j(\neq i)$ are negligible because $|g^{ij}|\ll |g^{ii}|$. On the other hand, because $p \ll \rho c^2$ in all main-sequence stars, the right-hand side (RHS) of Equation~\ref{eq:field} reduces to $-4\uppi G \rho$. Substituting the form for $g_{\mu\nu}$ given in Equation~\ref{eq:metric} into the t--t element of Equation~\ref{eq:field} then yields
\begin{align}\label{eq:R2}
     \frac{1}{2}\nabla^{2}h^{\rm tidal}_{tt} + \frac{1}{2}\nabla^{2}h^{\rm sg}_{tt} +4\uppi G \rho\simeq - \frac{1}{2}\sum_{i=1}^{3}g^{ii}(\partial_{t}^{2}g_{ii}-2\partial_{t}\partial_{i}g_{ti})-g^{\lambda\nu} g_{\eta\sigma}(\Gamma^{\eta}_{\lambda\nu}\Gamma^{\sigma}_{tt}-\Gamma^{\eta}_{t\lambda}\Gamma^{\sigma}_{t\nu}).
\end{align}
Because $R_{\mu\nu} = 0$ in the vacuum Schwarzschild spacetime, the sum of all terms in Equation~\ref{eq:R2} independent of $h^{\rm sg}_{tt}$ and $\rho$ must be zero. Thus, any relativistic corrections to the Poisson equation for $h_{tt}^{\rm sg}$ must be proportional to at least one factor of both $h_{tt}^{\rm sg}$ and $h_{\mu\nu}^{\rm tidal}$.

In view of the fact that $|h_{tt}^{\rm sg}| \ll |h_{\mu\nu}^{\rm tidal}| \lesssim 1$, it is convenient to consider only the leading-order terms, i.e. those proportional to $h^{\rm sg}_{tt} h_{\mu\nu}^{\rm tidal}$. 
Consider the first bracket on the RHS of Equation~\ref{eq:R2}.  Because both $g_{ii}$ and $g_{ti}$ are independent of $h_{tt}^{\rm sg}$ to lowest order, the only coupling to stellar self-gravity is through $g^{ii}$; although $g_{ii}$ contains no leading-order terms $\propto h_{tt}^{\rm sg}$, $g^{ii}$ can. The portion of $g^{ii}$ proportional to a single power of $h_{tt}^{\rm sg}$ is
\begin{align}
    2 h^{\rm sg}_{tt}(\sum_{j=1}^{3}h^{\rm tidal}_{jj}-h^{\rm tidal}_{ti}).
\end{align}
Consequently, the lowest-order corrections to the Poisson equation originating in the portion of the Ricci tensor linear in the connections are all second-order in $h_{\mu\nu}^{\rm tidal}$ and may be neglected.

The terms quadratic in the connections can be simplified in similar ways.  Any terms proportional to $h_{tt}^{\rm sg}$ due to its appearance in $g^{\lambda\nu}$ or $g_{\eta\sigma}$ are multiplied by two factors of metric gradients; each is $\propto h_{\mu\nu}^{\rm tidal}$, and is therefore second-order.  Consequently, the leading-order terms are those containing a product of one gradient of $h_{tt}^{\rm sg}$, one gradient of $h^{\rm tidal}$, and the Minkowski portion of $g^{\lambda\nu}g_{\eta\sigma}$:
\begin{align}
 \label{eq:leading2}
g^{\lambda\nu} g_{\eta\sigma}(\Gamma^{\eta}_{\lambda\nu}\Gamma^{\sigma}_{tt}-\Gamma^{\eta}_{t\lambda}\Gamma^{\sigma}_{t\nu})
  \simeq -\frac{1}{4}\sum_{i=1}^{3}[2\partial_{i}h^{\rm sg}_{tt}\partial_{i}h^{\rm tidal}_{tt}-\sum_{j=1}^{3}\partial_{j}h^{\rm sg}_{tt}(2\partial_{i}h^{\rm tidal}_{ij}-\partial_{j}h^{\rm tidal}_{ii})],
\end{align}
where the first terms in the square bracket on the RHS derive from $\eta=t$ and $\sigma=t$, and the remaining terms from $\eta=j$ and $\sigma=j$. 
For estimation purposes, we may therefore write equation~\ref{eq:R2} as
\begin{equation}\label{eq:R3}
\frac{1}{2}\nabla^{2}h^{\rm sg}_{tt} -  \beta \sum_{i=1}^{3}\partial_i h_{tt}^{\rm sg}\partial_{i} |h^{\rm tidal}| \simeq- 4\uppi G \rho ,
\end{equation}
where the $\beta \sim O(1)$, $|h^{\rm tidal}|$ is the typical magnitude of the tidal terms and $\partial_i$ is a stand-in for the appropriate spatial gradient.
Thus, Equation~\ref{eq:R3} may be regarded as a perturbed version of the Poisson equation, but one remaining linear in $h_{tt}^{\rm sg}$. These lowest-order relativistic corrections do not introduce any time-derivatives of $g_{tt}$, validating the ``snapshot" assumption. Our approximation is valid to the extent these perturbation terms have little effect on the solution.

Because the equation is linear in $h_{tt}^{\rm sg}$, the fractional error induced in the solution by neglect of the relativistic corrections is the same as the ratio of these corrections to the original terms. The relative error $D \sim |h^{\rm tidal}| R_{\star}/r_{\rm p}$ can then be estimated as
\begin{align}\label{eq:D}
   D \simeq 3~\beta ~\lambda~\left(\frac{R_{\star}}{r_{\rm g}}\right)\left(\frac{r_{\rm p}}{r_{\rm g}}\right)^{-1-n}
\end{align}
because the most relevant spatial scale is $r_{\rm p}$ and the gradient scale for the stellar potential is $R_{\star}$.  For the derivation of Equation~\ref{eq:D}, we have used the following scalings,
\begin{align}\label{eq:scaling1}
\partial_{i}&\simeq \left(\frac{r_{\rm p}}{r_{\rm g}}\right)^{-1}r_{\rm g}^{-1}\hspace{0.1in}\text{(for $h^{\rm tidal}_{\mu\nu}$)},\\\label{eq:scaling3}
\partial_{i}&\simeq \left(\frac{R_{\star}}{r_{\rm g}}\right)^{-1}r_{\rm g}^{-1}\hspace{0.07in}\text{(for $h^{\rm sg}_{tt}$)},\\\label{eq:scaling4}
\nabla^{2}&\simeq \left(\frac{R_{\star}}{r_{\rm g}}\right)^{-2}r_{\rm g}^{-2}.
\end{align}

It is instructive to see the dependence of $D$ on $M_{\rm BH}$ and $M_{\star}$. Replacing $r_{\rm p}$ with $\beta^{-1}~ r_{\rm t}$ in Equation~\ref{eq:D}, we find
\begin{align}\label{eq:D3}
   D&\simeq \lambda ~\beta^{(n+1)}~M_{\rm BH}^{(2n-1)/3}~M_{\star}^{(n+1)/3}~R_{\star}^{-n},\\\label{eq:D3}
 & \simeq \lambda ~\beta^{(n+1)} ~M_{\rm BH}^{0.66(n-1/2)}~M_{\star}^{-0.55(n-0.61)},\\\label{eq:D2}
  & \simeq 7\times10^{-5}~\beta^{2.6} \left(\frac{\lambda}{0.2}\right)\left(\frac{M_{\rm BH}}{10^{7}}\right)^{0.73}M_{\star}^{-0.55} \hspace{1.2in}\text{(for $n=1.6$)},
\end{align}
where we have used the $M_{\star}-R_{\star}$ relation that we find for our {\small MESA} models within $0.15\leq M_{\star}\leq3$, i.e., $R_{\star}=M_{\star}^{0.88}$ (Equation \ref{eq:M_R_relation}). Therefore, $D \lesssim O(10^{-4})$ for $M_{\rm BH}\lesssim10^{7}$ because $\Psi(M_{\rm BH}=10^7) \geq 0.65$ for $M_{\star} \lesssim 3$. 

Thus, granted the assumption that $|h^{\rm tidal}| \ll 1$, our Poisson equation solution for the stellar self-gravity should be quite accurate up to $M_{\rm BH} \simeq 3 \times 10^7$.   However, the actual limiting factor for our procedure is the validity of the assumption that the tidal perturbations are small; our estimate of the error depends upon this assumption's validity. As we have seen, when $r \lesssim 10~\rg$, $|h^{\rm tidal}|$ rises to $\gtrsim O(10^{-2})$; it is this that sets the limit on the applicability of our method.

Our results may be compared to those of \citet{ChengEvans2013}, who studied the tidal disruption of a white dwarf using relativistic hydrodynamic simulations. Their numerical methods for relativistic simulations are similar to ours in terms of self-gravity calculations: the star's self-gravity is calculated using a Newtonian Poisson solver in a frame co-moving with the star's center-of-mass.
However, their co-moving frame was defined in terms of Fermi normal coordinates rather than a tetrad system.  Consequently, their tidal terms had to be computed separately (by a multipole expansion), whereas ours are determined exactly by a coordinate transformation. In terms of their tidal terms (which we call $h_{CE}$), they estimated that $D \sim |h_{CE}|$, without reference to the different gradient lengthscales.  As a result, their estimated fractional error scales differently than our error estimate.

\begin{figure}
	\centering
	\includegraphics[width=8.3cm]{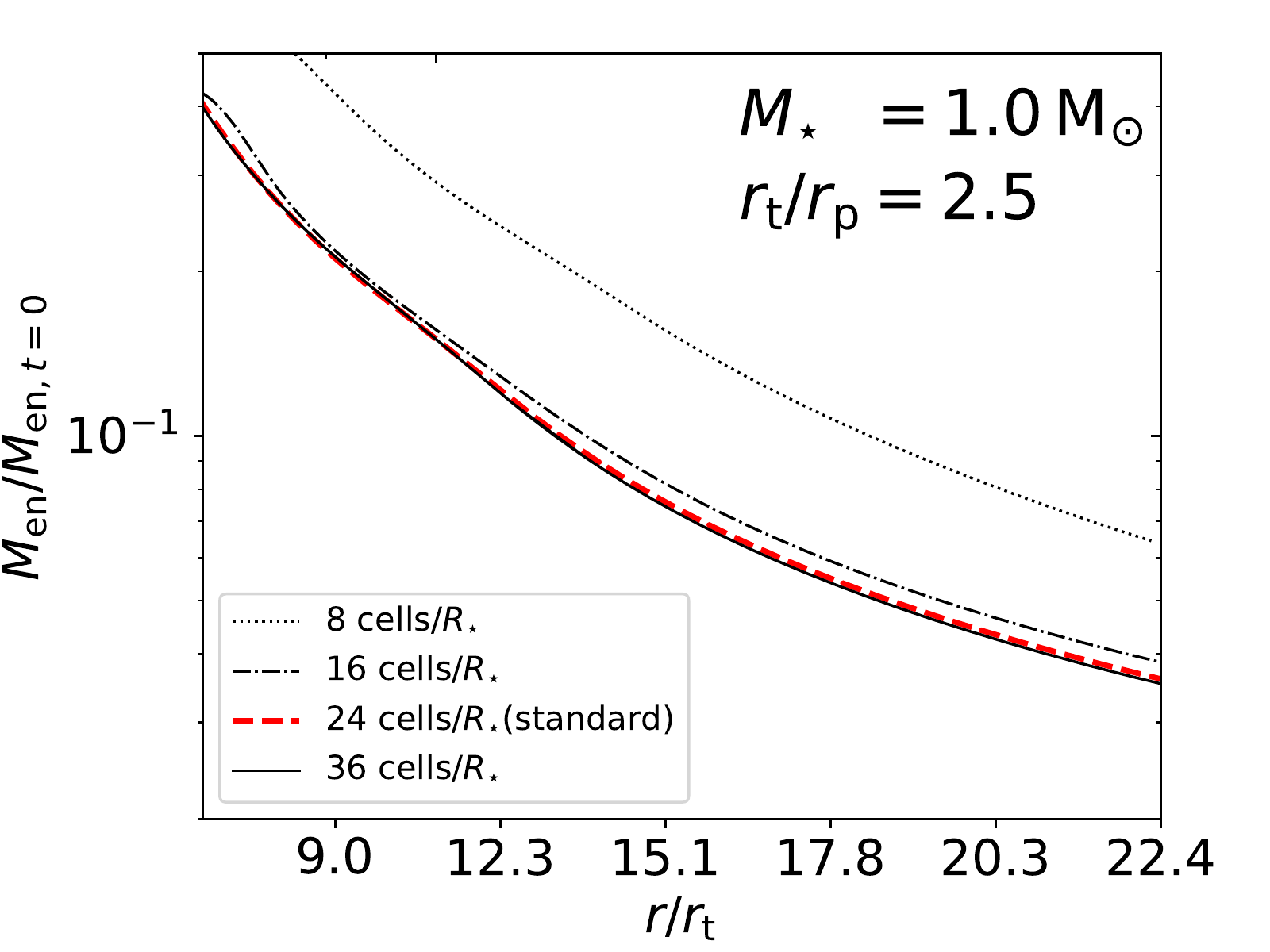}	
	\includegraphics[width=8.3cm]{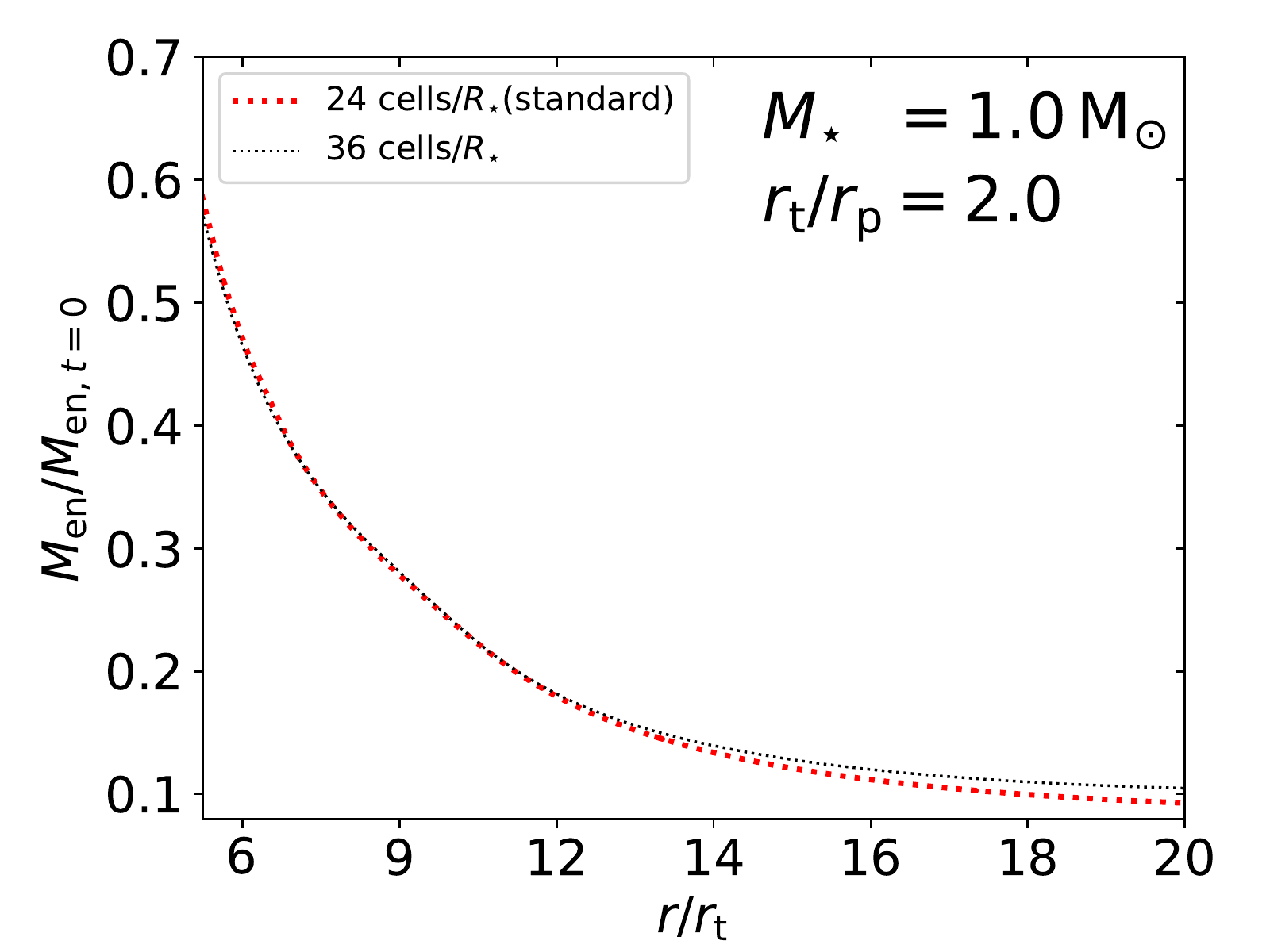}	
	\caption{The evolution, as the stellar debris move away from the SMBH, of mass enclosed in the computational domain $M_{\rm en}$, relative to its initial mass $M_{{\rm en},t=0}$,   for a full (the \textit{left} panel, $r_{\rm t}/r_{\rm p}=2.5$) and partial (the \textit{right} panel, $r_{\rm t}/r_{\rm p}=2.0$) TDE simulations ($M_{\star}=1$, $M_{\rm BH}=10^{6}$) with different resolutions. The red dashed line represents 
		the standard resolution of 24 cells per $R_{\star}$.}
	\label{fig:convergence}
\end{figure}

\begin{figure}
	\centering
	\includegraphics[width=9.1cm]{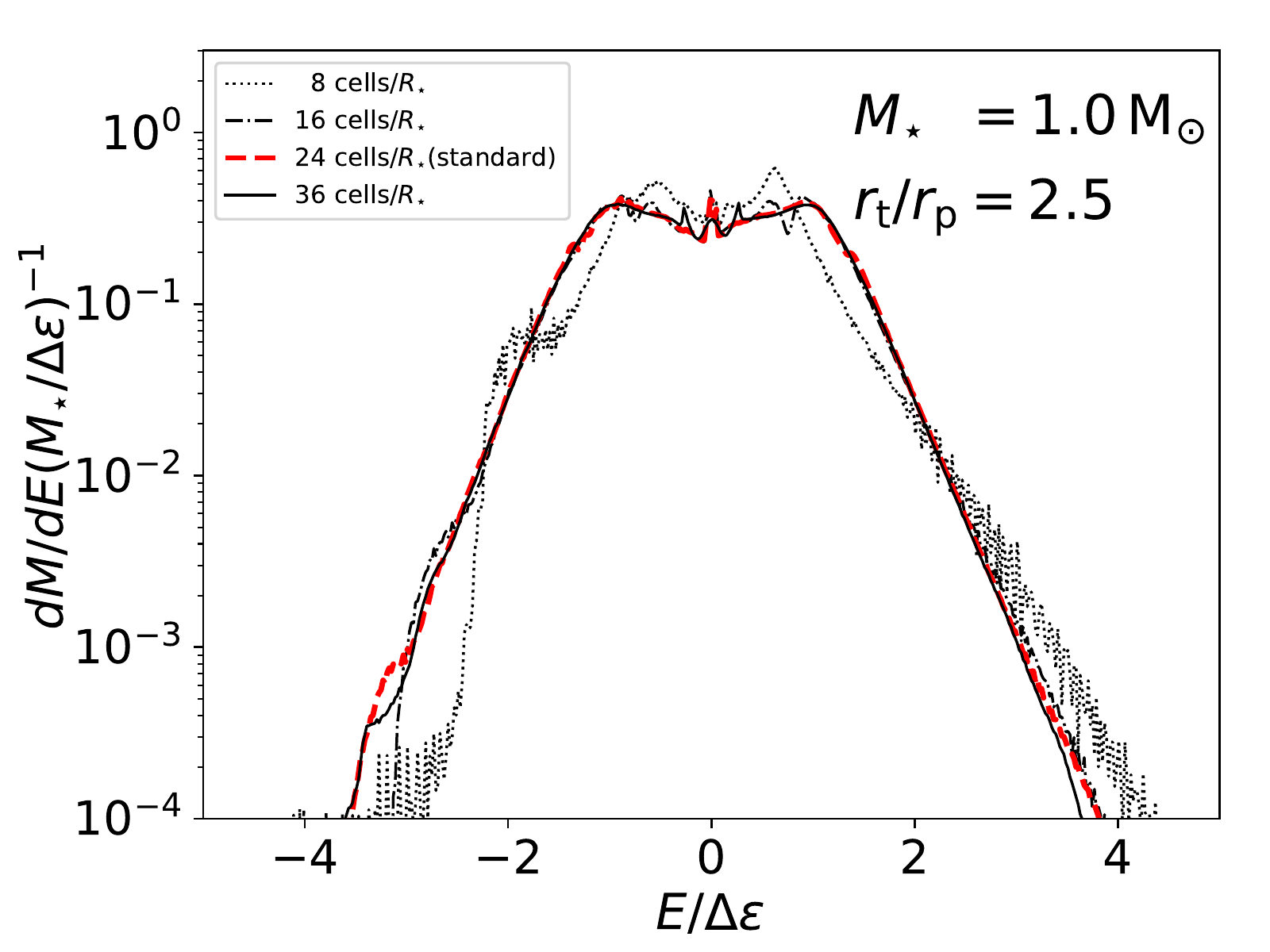}	
	\caption{The normalized energy distribution of the stellar debris produced in the same  full disruption TDE simulations ($r_{\rm t}/r_{\rm p}=2.5$) with  four different resolutions. The standard resolution (24 cells per $R_{\star}$) is marked in a red dashed line.}
	\label{fig:convergence2}
\end{figure}

\section{Convergence}\label{appendix2}

To show how well-converged our simulations are, we performed simulations with a number of different spatial resolutions for both a full ($r_{\rm t}/r_{\rm p}=2.5$) and a partial ($r_{\rm t}/r_{\rm p}=2.0$) tidal disruption of a $1\Msol$ star by a $10^{6}\Msol$ BH. For contrast with the resolution of our ``standard'' simulations (24 cells per $R_{\star}$), we considered grids with 8 cells/$R_{\star}$, 16 cells/$R_{\star}$, and 36 cells/$R_{\star}$.

In this appendix, we focus on two quantities whose convergence behavior indicates the accuracy of our calculations. Figure \ref{fig:convergence} shows the mass remaining in the computational domain $M_{\rm en}$ in ratio to the initial mass $M_{{\rm en},t=0}$ as a function of time for a full (\textit{left} panel) and a partial (\textit{right} panel) disruption. Note that the time evolution of this quantity is used as one of the criteria for full disruptions (see criteria \label{con3} in Section~\ref{subsec:disruptioncondition}). For partial disruptions, the late-time value of this quantity is the remnant mass (\citetalias{Ryu3+2019}).

For the full disruption case (the \textit{left} panel), the pace of mass-loss increases with finer resolution, but even with 16 cells/$R_{\star}$, the rate of mass-loss is within a few percent of the rate produced by a simulation in which each cell-dimension is another factor of 2 smaller.  There is almost no difference between the 24 cells/$R_{\star}$ curve and the 36 cells/$R_{\star}$ curve.  Thus, our standard resolution (24 cells/$R_{\star}$) is clearly well-converged with respect to this property.

The partial disruption case whose sensitivity to resolution we explore (the \textit{right} panel) is the most severe partial disruption we studied. Once again, the rate of mass-loss found with our standard resolution is very close to that given by a higher resolution run.  The fractional difference in remnant mass at $r\simeq 20~r_{\rm t}$  between the two resolutions is only $\simeq 3\%$.  The curves' slopes show that at this time the mass remaining is very nearly the asymptotic remnant mass.  We also confirmed that the error in $M_{\rm en}$ for a slightly less severe partial disruption ($r_{\rm t}/r_{\rm p}=1.82)$ is similar. Because weaker partial disruptions probe stellar layers at larger distance from the star's center, and the scale-length of internal density structure generically increases outward, we expect that the case we show displays the {\it greatest} departure in $M_{\rm en}$ from exact convergence.

In Figure \ref{fig:convergence2}, we depict the normalized energy distribution of the stellar debris in the full disruption case.  Simulating with only 8 cells per stellar radius produces a distribution noticeably different from that of higher resolution calculations, with sharp bends and noise features unlikely to be physical.   However, the global features of the distributions produced by any simulation with at least 16 cells per stellar radius are all very close to one another.  In particular, $\Delta E$, the quantity of greatest interest, is essentially identical in all three higher resolutions because it is defined as an integral: the energy-width containing 90\% of the bound mass.

For grids with more than 16 cells per stellar radius, the principal gain from finer resolution comes from more reliable determination of energy distribution features containing small amounts of mass, e.g., the wings of the distribution.  On the unbound side, the slope at progressively higher energies becomes slightly steeper with greater resolution; the difference in $dM/dE$ between 24 cells/$R_*$ and 36 cells/$R_*$ becomes $\gtrsim 10\%$ for values of normalized $dM/dE \lesssim 3 \times 10^{-4}$.  On the bound side, the predictions of these two runs differ at this level for normalized $dM/dE\lesssim 1 \times 10^{-3}$.

Lastly, we point out that in all cases there is a small feature at $E=0$.  This represents the mass remaining in the box at the end of the simulation, typically $\simeq 1-2\%$ (no more than $3-4\%$) of the initial stellar mass for our total disruption cases.  If the simulation were carried further, this bump would disappear.  There also remain some smaller irregularities in the range $|E/\Delta\epsilon| \lesssim 1$ that diminish with improved resolution.  None of these features, however, has any significant impact on the mass fallback rates shown in Figure~\ref{fig:dmde}, particularly at times when the fallback rate is great enough to be of observational interest.

\end{document}